\documentclass[%
aip,
amsmath,amssymb,
reprint, onecolumn 
]{revtex4-2}

\usepackage{xcolor}
\usepackage{graphicx}
\usepackage{dcolumn}
\usepackage{bm}

\usepackage[utf8]{inputenc}
\usepackage[T1]{fontenc}
\usepackage{mathptmx}
\usepackage{etoolbox}

\makeatletter
\def\@email#1#2{%
	\endgroup
	\patchcmd{\titleblock@produce}
	{\frontmatter@RRAPformat}
	{\frontmatter@RRAPformat{\produce@RRAP{*#1\href{mailto:#2}{#2}}}\frontmatter@RRAPformat}
	{}{}
}%
\makeatother
\begin{document}
	
	\preprint{AIP/123-QED}
	
	\title[Discrete Boltzmann model at Burnett level for
	compressible multicomponent flows under external
	forces]{Discrete Boltzmann model at Burnett level for
		compressible multicomponent flows under external
		forces}
	
	\author{Demei Li}
	\affiliation{School of Mathematics and Statistics \& Key Laboratory of Analytical Mathematics and Applications (Ministry of Education) , Fujian Normal University, 350117 Fuzhou, P. R. China}
	
	\author{Huilin Lai}
\affiliation{School of Mathematics and Statistics \& Key Laboratory of Analytical Mathematics and Applications (Ministry of Education) , Fujian Normal University, 350117 Fuzhou, P. R. China}
	
	\author{Chuandong Lin\footnote{linchd3@mail.sysu.edu.cn}}
	\affiliation{Sino-French Institute of Nuclear Engineering and Technology, Sun Yat-sen University, Zhuhai 519082, China}
	\affiliation{State Key Laboratory of Explosion Science and Safety Protection, Beijing Institute of Technology, Beijing 100081, China.}
	
	\author{Suni Chen}
	\affiliation{School of Mathematics and Statistics \& Key Laboratory of Analytical Mathematics and Applications (Ministry of Education) , Fujian Normal University, 350117 Fuzhou, P. R. China}
	
	\date{\today}

\begin{abstract}
This work extends the Burnett-level discrete Boltzmann model (DBM) from single-component to multicomponent compressible flows under external forces, building on the fundamental framework of the high-precision discrete kinetic method. A high-isotropy 25-discrete-velocity set is adopted to guarantee numerical stability and spatial symmetry, while a rigorous moment-matching strategy is developed to construct the equilibrium distribution function and external force term. Different from the single-component counterpart, the present model intrinsically incorporates interspecies mass diffusion effects and multi-component thermodynamic nonequilibrium behaviors, which are critical for complex compressible multicomponent systems. The Chapman--Enskog expansion verifies that the proposed model can exactly recover the Burnett-level governing equations for forced multicomponent compressible flows in the continuum limit. Five canonical benchmark cases, including multicomponent mass diffusion, compressible Sod shock tube, thermal Couette flow, Kelvin--Helmholtz instability, and Rayleigh--Taylor instability, are systematically performed. Numerical results demonstrate that the developed Burnett-level multicomponent DBM achieves high accuracy and robustness in capturing both hydrodynamic evolution and multicomponent nonequilibrium characteristics under external forces.
\end{abstract}

\maketitle

\section{Introduction}

Compressible multicomponent flows involve the simultaneous evolution of multiple fluid species, accompanied by spatiotemporally varying density, momentum, and energy fields as well as complex thermodynamic nonequilibrium (TNE) effects. Such flow phenomena are ubiquitous in natural processes and industrial applications, ranging from hypersonic aerodynamics and multispecies gas mixing to microscale transport and multiphase multicomponent systems~\cite{kundu2024fluid}. Differences in thermophysical properties among species, together with interspecies diffusion, heat transfer, and possible chemical reactions, jointly generate highly heterogeneous velocity, temperature, and concentration distributions. Compared with single-component flows, multicomponent compressible systems exhibit richer flow physics and more prominent nonequilibrium behaviors~\cite{Henry2026cpf,CUETOFELGUEROSO2025CMAME,guo2026CED}. Species-dependent variations in density, viscosity, and specific heat lead to distinct hydrodynamic responses under compression, expansion, and high-speed shear, while strongly coupled mass, momentum, and energy transport further complicate the flow dynamics. These inherent complexities pose considerable challenges for high-fidelity numerical modeling~\cite{Zhang2023JCP,Adebayo2025Fluids}. Accurate and robust numerical tools for compressible multicomponent nonequilibrium flows are therefore essential for advancing aerospace design, high-pressure energy transport, and multiscale multi-fluid flow research.

From the perspective of physical modeling, existing numerical approaches for compressible multispecies flows can be generally categorized into macroscopic, mesoscopic, and microscopic frameworks. Macroscopic methods based on the continuum hypothesis solve multicomponent Euler or Navier–Stokes equations using macroscopic field variables. Nevertheless, they rely on the local thermodynamic equilibrium assumption and tend to lose accuracy in regions with strong flow gradients, where multicomponent coupling and nonequilibrium effects dominate~\cite{PareteKoon2013}. In contrast, microscopic methods such as molecular dynamics can precisely resolve molecular interactions and TNE behaviors, but their prohibitive computational cost restricts their application to small-scale or short-time problems, limiting their feasibility for large-scale engineering simulations~\cite{Mao2023}.

As an intermediate bridge between macroscopic continuum solvers and microscopic molecular simulations, mesoscopic kinetic methods achieve a favorable balance between computational efficiency and physical fidelity. These methods naturally capture nonequilibrium flow characteristics and flexibly handle complex flow configurations, rendering them powerful for investigating shock structures, hydrodynamic instabilities, and multiphase mixing problems~\cite{DiIlio2021}. As one of the most successful mesoscopic kinetic approaches, the lattice Boltzmann method (LBM) describes fluid evolution via particle distribution functions on discrete lattices, rather than directly solving macroscopic governing equations~\cite{Succi2001,guo2013lattice,kruger2017lattice,gu2002PRE}. Over the past decades, the LBM has been extensively extended to turbulence, heat transfer, microfluidics, multiphase interfacial flows, mass transfer, and complex fluid-structure interaction problems~\cite{aidun2010lattice,wei2022small,liu2022lattice,wei2026pof,ZHANG2021APE,li2020multiphase,li2025POF,zhan2025PRE,liang2024pof,liang2026EC,chen2026IJMT,qiao2026JFM,zhao2026NHTA,wu2025CS,li2025lattice,yu2026JCP,HUANG2026JCP,chen2024JCP,Song2024POF,QIN2026AMM,qin2026JFM}. Despite these substantial progresses, conventional LBMs are primarily designed to recover Navier–Stokes-level macroscopic equations, which limits their performance in high-Mach-number compressible flows, where accurate resolution of shock structures, high-order energy transport, and strong TNE effects remains challenging~\cite{Gan2013EL}.

To overcome the inherent limitations of standard LBMs for compressible nonequilibrium flows, the discrete Boltzmann method (DBM) has been developed as a high-fidelity mesoscopic kinetic framework~\cite{xu2024FOP,ji2022CJP}. Unlike traditional LBMs that tightly couple time stepping, spatial grids, and discrete particle velocities, the DBM fully decouples these three components and thereby provides superior flexibility for modeling complex unsteady compressible flows~\cite{Lin2023CPB}. Benefiting from flexible discrete velocity configurations and hierarchical high-order kinetic moment analysis, the DBM is capable of quantitatively characterizing TNE behaviors and capturing fine flow structures in compressible and rarefied flow regimes~\cite{Shan2023CTP,Gan2019FOP}. In recent years, the DBM has developed into a versatile multiscale numerical tool for hydrodynamic instabilities, reactive combustion flows, multiphase systems, and plasma flows, with continuous advances in numerical schemes and physical modeling capabilities~\cite{Chen2022PRE,Li2022FOP,Lai2024CAF,xu2025POF,Su2022CTP,Li2022CTP,song2025pof,lin2026IJMPC,LIN2026CNSNS,ZHANG2023CAF}.

To improve the description of strong nonequilibrium effects, various high-order DBM models have been proposed in recent years. Gan et al. developed a BGK-type DBM for Burnett-level compressible flow simulations~\cite{Gan_2018pre}, while Zhang et al. constructed elliptical statistical BGK-based Burnett-level DBMs to resolve multi-order TNE effects under varying nonequilibrium intensities~\cite{Zhang_2018,Zhang_2022}. Gan et al. further established a systematic multiscale DBM framework bridging continuum and transitional flow regimes~\cite{Gan_2022}. Wang et al. proposed a high-order DBM with external force and interaction source terms for multiphase flow simulations based on Hermite polynomial expansion~\cite{Wang_2023}. Most recently, Chen et al. devised a Burnett-level DBM with a high-symmetry D2V25 discrete velocity set for forced single-component compressible flows, which exactly recovers Burnett-level equations and achieves accurate TNE characterization~\cite{chen2025pof}. Complementarily, Wu et al. developed MRT central-moment DBMs for chemically reactive compressible flows~\cite{wu2025caf}.

Despite these significant advances, most existing Burnett-level DBMs are restricted to single-component flow systems and cannot resolve species-specific thermohydrodynamic differences, including species segregation, differential diffusion, and multicomponent nonequilibrium coupling. Such limitations hinder their applications to multi-fluid interfacial instabilities, species mixing, and stratified compressible flows. To address this issue, several efforts have been devoted to developing multicomponent DBMs. Early multicomponent DBM variants were designed for reactive and premixed combustion flows~\cite{LIN2016,LIN2017PRE}. Zhang et al. proposed an ellipsoidal statistical BGK-based two-fluid DBM for compressible flows~\cite{Zhang2020}. Lin et al. developed MRT-type multicomponent DBMs with flexible heat ratio and Prandtl number adjustment, which are capable of simulating compressible mixing and hydrodynamic instabilities~\cite{Lin2021PRE}. Further extensions include split-collision MRT DBMs for reactive nonequilibrium flows and reaction-coupled MRT DBMs for multispecies transport~\cite{Lin2024CTP,Huang2025ATE}. These multicomponent DBMs have been widely applied to investigate RT, KH, and RM instabilities in multi-fluid systems~\cite{Lai2025POF,Li2024POF,lai2025compressible,Li2025TCFD,Lai2026JFM,Lian2025CTP,Lai2025RM}.

Nevertheless, existing multicomponent DBMs are mostly limited to Navier--Stokes-level accuracy, while high-order Burnett-level multicomponent DBMs that simultaneously account for external forces, interspecies mass diffusion, and multicomponent TNE coupling remain absent. This research gap motivates the present work. Building on our previous single-component Burnett-level DBM for forced compressible flows~\cite{chen2025pof}, this study develops a novel Burnett-level discrete Boltzmann model for compressible multicomponent flows under external forces. The adopted high-symmetry D2V25 discrete velocity structure guarantees excellent numerical isotropy and stability. The present model inherently incorporates interspecies mass diffusion and multicomponent coupled nonequilibrium effects, enabling high-order kinetic modeling of forced compressible multispecies flows. A suite of canonical benchmarks, including multicomponent diffusion, Sod shock tube, thermal Couette flow, Kelvin--Helmholtz instability, and Rayleigh--Taylor instability, are conducted to validate the accuracy, robustness, and physical reliability of the proposed model.

The remainder of this paper is organized as follows. Section~\ref{sec:modeling} elaborates the detailed construction and theoretical derivation of the proposed multicomponent Burnett-level DBM. Section~\ref{sec:Verification and validation} presents systematic numerical validations and physical discussions. Finally, Section~\ref{sec:conclusion} summarizes the main conclusions and prospects future research directions.

\section{Discrete Boltzmann model}\label{sec:modeling}

\subsection{Discrete Boltzmann equation}
The discrete Boltzmann equation for multicomponent flows can be formulated as~\cite{LIN2016}:
\begin{equation}
	\frac{\partial {{f}_{i}^{\sigma}}}{\partial t}+{{\mathbf{v}}_{i}^{\sigma}}\cdot \nabla {{f}_{i}^{\sigma}}=-\frac{1}{\tau^{\sigma} }\left( {{f}_{i}^{\sigma}}-f_{i}^{\sigma\rm{eq}} \right)+{F_i^{\sigma}}
	\label{DBEquation}\text{,}
\end{equation}
where $\sigma$ represents chemical species, $t$ the time, $\tau^{\sigma} $ the relaxation time, $\nabla$ the nabla operator,  ${f}_{i}^{\sigma}$ ($f_{i}^{\sigma\rm{eq}}$) the discrete (equilibrium) distribution function, and $i$ ($=1$, $2$, $\cdots$, $N$) the index of the discrete velocities ${\mathbf{v}}_{i}^{\sigma} $, with $N$ being the total number of discrete velocities. In the right-hand term of the aforementioned equation, ${F}_{i}^{\sigma}$ indicates the force term, which is calculated by Eq. (\ref{Expression_force_i}).

Constructing a Burnett-level DBM requires its discrete equilibrium distribution function to satisfy nine prescribed sets of kinetic moment relations.
\begin{equation}
	\sum\nolimits_{i}{f_{i}^{\sigma\rm{eq}}}=\iint{{{f}^{\sigma\rm{eq}}}d\mathbf{v}d\eta }
	\label{Moment_feq0}
	\text{,}
\end{equation}
\begin{equation}
	\sum\nolimits_{i}{f_{i}^{\sigma\rm{eq}}{{v}_{i\alpha }^{\sigma}}}=\iint{{{f}^{\sigma\rm{eq}}}{{v}_{\alpha }}d\mathbf{v}d\eta }
	\label{Moment_feq1}
	\text{,}
\end{equation}
\begin{equation}
	\sum\nolimits_{i}{f_{i}^{\sigma\rm{eq}}\left( v_{i}^{\sigma 2}+\eta _{i}^{\sigma 2} \right)}=\iint{{{f}^{\sigma\rm{eq}}}\left( {{v}^{2}}+{{\eta }^{2}} \right)d\mathbf{v}d\eta }
	\label{Moment_feq2,0}
	\text{,}
\end{equation}
\begin{equation}
	\sum\nolimits_{i}{f_{i}^{\sigma\rm{eq}}{{v}_{i\alpha }^{\sigma}}{{v}_{i\beta }^{\sigma}}}=\iint{{{f}^{\sigma\rm{eq}}}{{v}_{\alpha }}{{v}_{\beta }}d\mathbf{v}d\eta }
	\label{Moment_feq2}
	\text{,}
\end{equation}
\begin{equation}
	\sum\nolimits_{i}{f_{i}^{\sigma\rm{eq}}\left( v_{i}^{\sigma 2}+\eta _{i}^{\sigma 2} \right)v_{i\alpha }^{\sigma }}=\iint{{{f}^{\sigma \rm{eq}}}\left( {{v}^{ 2}}+{{\eta }^{ 2}} \right){{v}_{\alpha }}d\mathbf{v}d\eta }
	\label{Moment_feq3,1}
	\text{,}
\end{equation}
\begin{equation}
	\sum\nolimits_{i}{f_{i}^{\sigma\rm{eq}}{{v}_{i\alpha }^{\sigma}}{{v}_{i\beta }^{\sigma}}{{v}_{i\chi }^{\sigma }}}=\iint{{{f}^{\sigma\rm{eq}}}{{v}_{\alpha }}{{v}_{\beta }}{{v}_{\chi }}d\mathbf{v}d\eta }
	\label{Moment_feq3}
	\text{,}
\end{equation}
\begin{equation}
	\sum\nolimits_{i}{f_{i}^{\sigma\rm{eq}}\left( v_{i}^{\sigma 2}+\eta _{i}^{\sigma 2} \right){{v}_{i\alpha }^{\sigma}}{{v}_{i\beta }^{\sigma}}}=\iint{{{f}^{\sigma \rm{eq}}}\left( {{v}^{ 2}}+{{\eta }^{ 2}} \right){{v}_{\alpha }}{{v}_{\beta }}d\mathbf{v}d\eta }
	\label{Moment_feq4,2}
	\text{,}
\end{equation}
\begin{equation}
	\sum\nolimits_{i}{f_{i}^{\sigma\rm{eq}}{{v}_{i\alpha }^{\sigma}}{{v}_{i\beta }^{\sigma}}{{v}_{i\chi }^{\sigma }}{{v}_{i\gamma }^{\sigma }}}=\iint{{{f}^{\sigma \rm{eq}}}{{v}_{\alpha }}{{v}_{\beta }}{{v}_{\chi }}{{v}_{\gamma }}d\mathbf{v}d\eta }
	\label{Moment_feq4}
	\text{,}
\end{equation}
\begin{equation}
	\sum\nolimits_{i}{f_{i}^{\sigma\rm{eq}}\left( v_{i}^{\sigma 2}+\eta _{i}^{\sigma 2} \right){{v}_{i\alpha }^{\sigma}}{{v}_{i\beta }^{\sigma}}{{v}_{i\chi }^{\sigma }}}=\iint{{{f}^{\sigma {eq}}}\left( {{v}^{ 2}}+{{\eta }^{ 2}} \right){{v}_{\alpha }}{{v}_{\beta }}{{v}_{\chi }}d\mathbf{v}d\eta }
	\label{Moment_feq5,3}
	\text{,}
\end{equation}
where $\eta_{i}^{\sigma}$ ($\eta_{i}$) is the (discrete) free parameter, $v=|\mathbf{v}|$ and ${v_{i}^{\sigma}}=|\mathbf{v}_{i}^{\sigma}|$ denote the magnitudes of the continuous and discrete particle velocities, respectively, while ${v}_{\alpha}^{\sigma}$ and ${v}_{i\alpha}^{\sigma}$ are their components in the $\alpha$-direction. In the two-dimensional case, the indices $\alpha$, $\beta$, $\chi$, and $\gamma$ each take values of $x$ or $y$. Additionally, the Maxwellian equilibrium distribution function ${{f}^{\sigma\rm{eq}}}$ takes the form~\cite{LIN2017PRE}:
\begin{equation}
	{{f}^{\sigma\rm{eq}}}=n^{\sigma}{{\left( \frac{m^{\sigma}}{2\pi kT} \right)}^{D/2}}{{\left( \frac{m^{\sigma}}{2\pi I^{\sigma}kT} \right)}^{1/2}}\exp \left[ -\frac{m^{\sigma}{{\left| \mathbf{v}-\mathbf{u} \right|}^{2}}}{2kT}-\frac{m^{\sigma}{{\eta }^{2}}}{2I^{\sigma}kT} \right]
	\text{,}
	\label{Expression_feq}
\end{equation}
where $m^{\sigma}$ stands for the molar mass, ${\mathbf{u}}$ the flow velocity, $T$ the mixture temperature, $n^{\sigma}$ the molar concentration,  ${\mathbf{v}}$ the particle velocity, $D = 2$ the spatial dimension, $k=1$ the Boltzmann constant, $I^{\sigma}$ the extra degrees of freedom corresponding to molecular vibration and/or rotation, and the symbol $\eta$ stands for the description of the extra internal energy.

Moreover, substituting Eq. (\ref{Expression_feq}) into Eqs. (\ref{Moment_feq0})-(\ref{Moment_feq5,3}) leads to a compact matrix form,
\begin{equation}
	{\mathbf{C}} \cdot {\mathbf{{f}^{\sigma\rm{eq}}}} = {\mathbf{\hat f^{\sigma\rm{eq}}}}
	\label{Matrix_form_feq}
	\text{,}
\end{equation}
where ${\mathbf{{f}^{\sigma\rm{eq}}}}=[f_{1}^{\sigma\rm{eq}},f_{2}^{\sigma\rm{eq}},\dots,f_{N}^{\sigma\rm{eq}}]^{\rm{T}}$ represents the column matrix formed by discrete equilibrium distribution functions; ${\mathbf{{\hat f}^{\sigma\rm{eq}}}}=[\hat f_{1}^{\sigma\rm{eq}}, \hat f_{2}^{\sigma\rm{eq}}, \dots, \hat f_{N}^{\sigma\rm{eq}}]^{\rm{T}}$ is the column matrix  equilibrium kinetic moments (refer to \ref{A}). Since in the two-dimensional case, Eqs. (\ref{Moment_feq0})-(\ref{Moment_feq5,3}) contain $N=25$ independent kinetic moments. ${\mathbf{C}}$ refers to a square matrix where all entries depends on the discrete parameters ${\mathbf{v}}_{i}^{\sigma}$ and ${\eta}_{i}^{\sigma}$ (refer to \ref{A}).
When ${\mathbf{C}}$ is invertible, Eq. (\ref{Matrix_form_feq}) can be expressed as~\citep{Lin2019PRE}:
\begin{equation}
	{\mathbf{{f}^{\sigma\rm{eq}}}} = {\mathbf{C}}^{-1} \cdot {\mathbf{{{\hat f}}^{\sigma\rm{eq}}}}
	\label{Expression_feq_i}
	\text{,}
\end{equation}
where ${\mathbf{C}}^{-1}$ denotes the inverse matrix associated with ${\mathbf{C}}$. Note that Eq. (\ref{Expression_feq_i}) provides the way how the discrete equilibrium distribution function is calculated.

Similarly, the force term ${F}_{i}^{\sigma }$ should also satisfy the following nine sets of kinetic moment relations:
\begin{equation}
	\sum\nolimits_{i}{F_{i}^{\sigma }}=\iint{{{F}^{\sigma }}d\mathbf{v}d\eta }
	\label{Moment_force0}
	\text{,}
\end{equation}
\begin{equation}
	\sum\nolimits_{i}{F_{i}^{\sigma }{{v}_{i\alpha }^{\sigma}}}=\iint{{{F}^{\sigma }}{{v}_{\alpha }}d\mathbf{v}d\eta }
	\label{Moment_force1}
	\text{,}
\end{equation}
\begin{equation}
	\sum\nolimits_{i}{F_{i}^{\sigma }\left( v_{i}^{\sigma 2}+\eta _{i}^{\sigma 2} \right)}=\iint{{{F}^{\sigma }}\left( {{v}^{ 2}}+{{\eta }^{2}} \right)d\mathbf{v}d\eta }
	\label{Moment_force2,0}
	\text{,}
\end{equation}
\begin{equation}
	\sum\nolimits_{i}{F_{i}^{\sigma }{{v}_{i\alpha }^{\sigma}}{{v}_{i\beta }^{\sigma}}}=\iint{{{F}^{\sigma }}{{v}_{\alpha }}{{v}_{\beta }}d\mathbf{v}d\eta }
	\label{Moment_force2}
	\text{,}
\end{equation}
\begin{equation}
	\sum\nolimits_{i}{F_{i}^{\sigma }\left( v_{i}^{\sigma 2}+\eta _{i}^{\sigma 2} \right)v_{i\alpha }^{\sigma }}=\iint{{{F}^{\sigma }}\left( {{v}^{ 2}}+{{\eta }^{2}} \right){{v}_{\alpha }}d\mathbf{v}d\eta }
	\label{Moment_force3,1}
	\text{,}
\end{equation}
\begin{equation}
	\sum\nolimits_{i}{F_{i}^{\sigma }{{v}_{i\alpha }^{\sigma}}{{v}_{i\beta }^{\sigma}}{{v}_{i\chi }^{\sigma }}}=\iint{{{F}^{\sigma }}{{v}_{\alpha }}{{v}_{\beta }}{{v}_{\chi }}d\mathbf{v}d\eta }
	\label{Moment_force3}
	\text{,}
\end{equation}
\begin{equation}
	\sum\nolimits_{i}{F_{i}^{\sigma }\left( v_{i}^{\sigma 2}+\eta _{i}^{\sigma 2} \right){{v}_{i\alpha }^{\sigma}}{{v}_{i\beta }^{\sigma}}}=\iint{{{F}^{\sigma }}\left( {{v}^{2}}+{{\eta }^{2}} \right){{v}_{\alpha }}{{v}_{\beta }}d\mathbf{v}d\eta }
	\label{Moment_force4,2}
	\text{,}
\end{equation}
\begin{equation}
	\sum\nolimits_{i}{F_{i}^{\sigma }{{v}_{i\alpha }^{\sigma}}{{v}_{i\beta }^{\sigma}}{{v}_{i\chi }^{\sigma }}{{v}_{i\gamma }^{\sigma }}}=\iint{{{F}^{\sigma }}{{v}_{\alpha }}{{v}_{\beta }}{{v}_{\chi }}{{v}_{\gamma }}d\mathbf{v}d\eta }
	\label{Moment_force4}
	\text{,}
\end{equation}
\begin{equation}
	\sum\nolimits_{i}{F_{i}^{\sigma }\left( v_{i}^{\sigma 2}+\eta _{i}^{\sigma 2} \right){{v}_{i\alpha }^{\sigma}}{{v}_{i\beta }^{\sigma}}{{v}_{i\chi }^{\sigma }}}=\iint{{{F}^{\sigma }}\left( {{v}^{ 2}}+{{\eta }^{ 2}} \right){{v}_{\alpha }}{{v}_{\beta }}{{v}_{\chi }}d\mathbf{v}d\eta }
	\label{Moment_force5,3}
	\text{,}
\end{equation}
where
\begin{equation}
	{F^{\sigma}}\approx-\frac{m^{\sigma}\mathbf{a}\cdot(\mathbf{v}-\mathbf{u^{\sigma}})}Tf^{\sigma\rm{eq}}
	\label{Expression_force_term}
	\text{,}
\end{equation}
which characterizes the variation rate of the distribution function induced by external forces~\cite{LIN2017PRE}. The nine sets of kinetic moment relations are formulated within a matrix framework~\cite{Lin2019PRE}:
\begin{equation}
	{\mathbf{C}} \cdot {\mathbf{F^{\sigma}}} = {\mathbf{\hat F^{\sigma}}}
	\label{Matrix_form_force}
	\text{,}
\end{equation}
where ${\mathbf F^{\sigma} }=[F_{1}^{\sigma} , F_{2}^{\sigma} ,\dots,F_{N}^{\sigma}]^{\rm{T}}$ refers to the column matrix of discrete velocity-space force components, ${\mathbf{{\hat F^{\sigma}}}}=[\hat F_{1}^{\sigma},\hat F_{2}^{\sigma},\dots,\hat F_{N}^{\sigma}]^{\rm{T}}$ represents the column matrix for continuous space force items (refer to \ref{A}).
Similarly, Eq. (\ref{Matrix_form_force}) can be expressed as:
\begin{equation}
	{\mathbf{F^{\sigma }}} = {\mathbf{C}}^{-1} \cdot {\mathbf{{{\hat F^{\sigma }}}}}
	\label{Expression_force_i}
	\text{,}
\end{equation}
when ${\mathbf{C}}$ is invertible. It should be pointed out that Eq. (\ref{Expression_force_i}) offers the expression of the force term.

Furthermore, the molar concentration of component $\sigma$, $n^{\sigma}$, is given by
\begin{equation}
	n^{\sigma} = \sum\nolimits_{i}f_{i}^{\sigma}
	\label{molar_concentration}
\end{equation}
The mass density of component $\sigma$, $\rho^{\sigma}$, reads
\begin{equation}
	\rho^{\sigma} = m^{\sigma} n^{\sigma}
	\label{Density}
\end{equation}
The flow velocity of component $\sigma$, $\mathbf{u}^{\sigma}$, is expressed as
\begin{equation}
	\mathbf{u}^{\sigma} = \dfrac{\sum\nolimits_{i}f_{i}^{\sigma} \mathbf{v}_{i}^{\sigma}}{n^{\sigma}}
\end{equation}
The internal energy density of component $\sigma$, $E^{\sigma}$, takes the form
\begin{equation}
	E^{\sigma} = \dfrac{m^{\sigma}}{2}\sum\nolimits_{i}f_{i}^{\sigma} \left( |\mathbf{v}_{i}^{\sigma}|^{2}+(\eta_{i}^{\sigma})^{2} \right)
	\label{Energy}
\end{equation}
The temperature of component $\sigma$, $T^{\sigma}$, is formulated as
\begin{equation}
	T^{\sigma}=\dfrac{2E^{\sigma}-\rho^{\sigma} |\mathbf{u}^{\sigma}|^{2}}{\left( D+I^{\sigma} \right)n^{\sigma}}
	\label{Temperature}
\end{equation} It is clear that the macroscopic quantities evolve as the $f_i^{\sigma }$ are updated in the process of the discrete Boltzmann equation.

The mixing number molar concentration $n$ is defined as
\begin{equation}
	n= \sum\nolimits_{\sigma}n^{\sigma }
	\label{molar_concentsrationsb}
	\text{.}
\end{equation}
The mixture mass density $\rho$ is given by
\begin{equation}
	\rho= \sum\nolimits_{\sigma}\rho^{\sigma }
	\label{molar_concentrationsb}
	\text{.}
\end{equation}
The flow velocity $\mathbf{u}$ reads
\begin{equation}
	\mathbf{u}= \dfrac{\sum\nolimits_{\sigma}\rho^{\sigma }\mathbf{u^{\sigma }}}{\rho}
	\label{molar_concentratiosnb}
	\text{.}
\end{equation}
The total energy density of the mixture $E$ is expressed as
\begin{equation}
	E= \sum\nolimits_{\sigma}E^{\sigma }
	\label{molar_concentratdionb}
	\text{.}
\end{equation}
The internal energy density $E_{\rm{int}}$ takes the form
\begin{equation}
	E_{\rm{int}}=E-\dfrac{1}{2}\rho \left|\mathbf{u}   \right|^2
	\label{molar_concentratdion}
	\text{.}
\end{equation}
The mixture temperature $T$ is formulated as
\begin{equation}
	T= \dfrac{2{{	E_{\rm{int}}}}}{\sum\nolimits_{\sigma}\left( D+{{I^{\sigma }}} \right){{n^{\sigma }}}}
	\label{molar_concenTGTHtratdionb}
	\text{.}
\end{equation}

\subsection{Nonequilibrium effects}

Apart from reproducing the conservation equations under  the continuum limit, the DBM is capable of extracting essential nonequilibrium information that goes beyond these equations. Specifically, if the discrete equilibrium  distribution functions ${f_i^{\sigma\rm{eq}}}$ are substituted with the discrete distribution functions ${f}_i^{\sigma}$, Eqs. (\ref{Moment_feq0}) - (\ref{Moment_feq2,0}) remain valid, which individually describe the conservation laws of mass, momentum, and energy. However, in a TNE state, substituting ${f_i^{\sigma\rm{eq}}}$ for ${f_i^{\sigma}}$ might break the balance of  Eqs. (\ref{Moment_feq2}) - (\ref{Moment_feq5,3}). Discrepancies in high-order kinetic moments solved from ${f_i^{\sigma}}$ and equilibrium distribution functions ${f_i^{\rm{eq}}}$ essentially quantify how significantly the physical system drifts away from local equilibrium. From a mathematical perspective, the forms describing nonequilibrium features are given below:
\begin{equation}
	\Delta^{\sigma*}_{2, \alpha \beta }=\sum\nolimits_{i}{f_{i}^{\sigma\text{neq}}{{v}^{\sigma*}_{i\alpha }}{{v}^{\sigma*}_{i\beta }}}
	\label{Delta2}
	\text{,}
\end{equation}
\begin{equation}
	\Delta^{\sigma*}_{3,1, \alpha }=
	\sum\nolimits_{i}{f_{i}^{\sigma\text{neq}}\left( v_{i}^{\sigma *2}+\eta _{i}^{\sigma 2*} \right)v_{i\alpha }^{\sigma}}
	\label{Delta31}
	\text{,}
\end{equation}
\begin{equation}
	\Delta^{\sigma*}_{3,\alpha \beta \chi}=
	\sum\nolimits_{i}{f_{i}^{\sigma\text{neq}}{{v}^{\sigma*}_{i\alpha }}{{v}^{\sigma*}_{i\beta }}{{v}^{\sigma*}_{i\chi }}}
	\label{Delta3}
	\text{,}
\end{equation}
\begin{equation}
	\Delta^{\sigma*}_{4,2, \alpha \beta}=
	\sum\nolimits_{i}{f_{i}^{\sigma\text{neq}}\left( v_{i}^{\sigma *2}+\eta _{i}^{\sigma* 2} \right){{v}^{\sigma*}_{i\alpha }}{{v}^{\sigma*}_{i\beta }}}
	\label{Delta42}
	\text{,}
\end{equation}
\begin{equation}
	\Delta^{\sigma*}_{4, \alpha \beta \chi \gamma}=
	\sum\nolimits_{i}{f_{i}^{\sigma\text{neq}}{{\left( v_{i}^{\sigma 2*}+\eta _{i}^{\sigma *2} \right)}}{{v}^{\sigma}_{i\alpha }}v_{i\beta }^{\sigma*}v_{i\chi }^{\sigma*}v_{i\gamma }^{\sigma*}}
	\label{Delta4}
	\text{,}
\end{equation}
\begin{equation}
	\Delta^{\sigma*}_{5,3, \alpha \beta \chi }=
	\sum\nolimits_{i}{f_{i}^{\sigma\text{neq}}{{\left( v_{i}^{\sigma 2*}+\eta _{i}^{\sigma 2*} \right)}}v_{i\alpha }^{\sigma*}v_{i\beta }^{\sigma*}v_{i\chi }^{\sigma}}
	\label{Delta53}
	\text{,}
\end{equation}
where \( f_{i}^{\sigma\mathrm{neq}} = f_{i}^{\sigma} - f_{i}^{\sigma\mathrm{eq}} \) denotes the nonequilibrium part of the discrete distribution function, and $v^{\sigma*}_{i\alpha}=v_{i\alpha}-u_{\alpha}$ represents the peculiar velocity. The symbols \( \Delta^{\sigma*}_{m,n} \) are the corresponding nonequilibrium moments, where the indices \( m \) and \( n \) denote the tensor contraction from rank \( m \) to rank \( n \). Physically, \( \Delta^{\sigma*}_{2,\alpha\beta} \) characterize the disordered momentum flux (i.e., the viscous effects), while \( \Delta^{\sigma*}_{3,1,\alpha} \) and \( \Delta^{\sigma*}_{3,\alpha\beta\chi} \) stand for unordered energy flux (i.e., heat conduction). The term \( \Delta^{\sigma*}_{4,2,\alpha\beta} \) characterizes the flux of irregular kinetic energy. The moments \( \Delta^{\sigma*}_{4,\alpha\beta\chi\gamma} \) and \( \Delta^{\sigma*}_{5,3,\alpha\beta\chi} \) account for higher-order nonequilibrium effects.

\subsection{Discretization of velocity}
It is widely recognized that cutting down the discrete velocity count helps reduce computational consumption and elevate computational performance. For the system of equations to be solvable, the number of discrete velocities must be at least equal to the number of independent kinetic moments required. The minimal configuration, where these two numbers are equal, can be obtained by the matrix match approach~\cite{Lin2019PRE}. Actually, in two dimensions, each of the nine sets of kinetic moment relations in Eqs.~(\ref{Moment_feq0}) - (\ref{Moment_feq5,3}) and Eqs.~(\ref{Moment_force0}) - (\ref{Moment_force5,3}) contains $25$ independent components. Consequently, The present work employs a total of $N = 25$ discrete velocity directions. Each discrete velocity direction corresponds to one discrete Boltzmann equation that needs to be solved, which makes the matrix mapping strategy the most computationally efficient scheme for DBM.
\begin{figure}
	\begin{center}
		\includegraphics[width=0.45\textwidth]{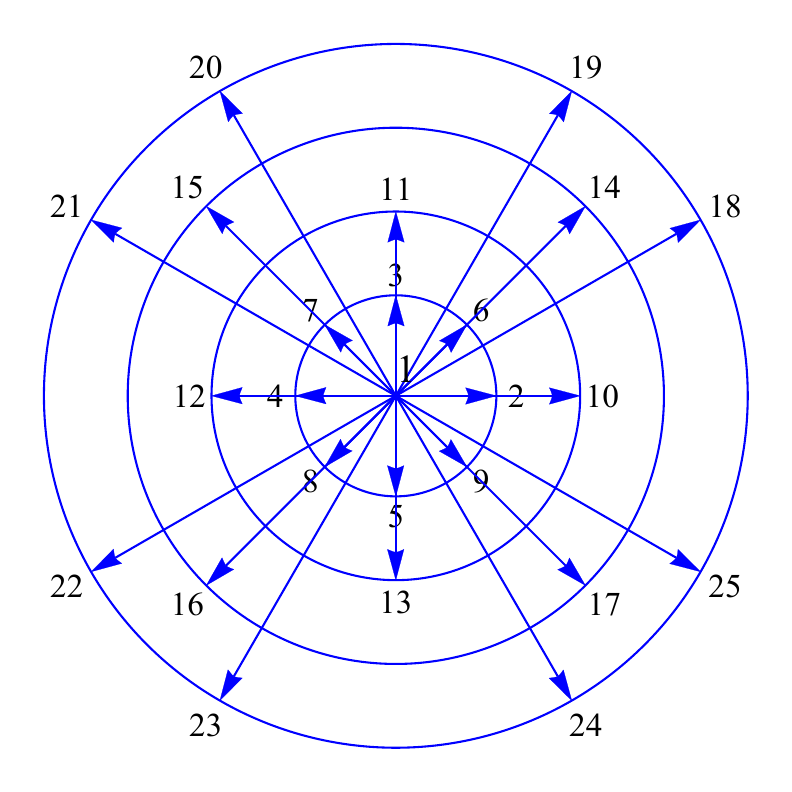}
	\end{center}
	\caption{Sketches of discrete velocities.}
	\label{Fig01}
\end{figure}

Moreover, numerical stability typically imposes more stringent requirements on the symmetry of the discrete velocity set. Good spatial symmetry helps suppress unphysical anisotropy caused by discretization errors, thereby ensuring a stable and reliable evolution. Therefore, to enhance numerical stability, it is often essential to construct a discrete velocity set with high symmetry, even if it exceeds the minimal requirement. Now, let us construct the discrete velocities, D2V25, with symmetry characteristics. As delineated in Fig. \ref{Fig01}, the discrete velocities take the form,
\begin{equation}
	\mathbf{v}_{i}^{\sigma} =
	\left\{
	\begin{array}{ll}
		0, & i = 1, \\ [10pt]
		\text{cyc}: v^{\sigma}_a \left(\pm 1, 0\right), & 2 \leq i \leq 5, \\ [10pt]
		v^{\sigma}_a \left(\pm \dfrac{\sqrt{2}}{2}, \pm \dfrac{\sqrt{2}}{2}\right), & 6 \leq i \leq 9, \\ [10pt]
		\text{cyc}: v^{\sigma}_b \left(\pm 1, 0\right), & 10 \leq i \leq 13, \\ [10pt]
		v^{\sigma}_c \left(\pm \dfrac{\sqrt{2}}{2}, \pm \dfrac{\sqrt{2}}{2}\right), & 14 \leq i \leq 17, \\ [10pt]
		\text{cyc}: v^{\sigma}_d \left(\pm \dfrac{\sqrt{3}}{2} , \pm 0.5\right), & 18 \leq i \leq 25 \text{,}
	\end{array}
	\right.
	\label{DVM_D2V25}
\end{equation}
where ${v}^{\sigma}_{a}$, ${v}^{\sigma}_{b}$, ${v}^{\sigma}_{c}$, and ${v}^{\sigma}_{d}$ are adjustable parameters that control the amplitude of each discrete velocity vector. Additionally, the parameter assignments for ${\eta}^{\sigma}_{i}$ read ${\eta}^{\sigma}_{1} = 0$, ${\eta}^{\sigma}_{2 \le i \le 13}={\eta}^{\sigma}_{a}$, and ${\eta}^{\sigma}_{14 \le i \le 25}={\eta}^{\sigma}_{b}$. The values of ${ \eta }^{\sigma}_{a}$ and ${\eta }^{\sigma}_{b}$ are also tunable.

It should be stressed that the values of $|\mathbf{v}_{i}^{\sigma}|$ and ${\eta }^{\sigma}_{i}$ can be adjusted to enhance the numerical stability and computational precision of the DBM. The magnitudes of ${{{\mathit{v}}}_{i}}$ can be set near the values of flow velocity magnitude $u^{\sigma} = |\mathbf{u}^{\sigma}|$ and acoustic velocity $v^{\sigma}_s = \sqrt{\gamma^{\sigma}{T^{\sigma}}/{m^{\sigma}}}$, with the specific heat ratio $\gamma^{\sigma}=(D+I^{\sigma}+2)/(D+I^{\sigma})$. Furthermore, the values of ${\eta}^{\sigma}_{i}$ should be distributed on both sides of the characteristic value $\bar{\eta}^{\sigma} = \sqrt{I^{\sigma} T^{\sigma} / m^{\sigma}}$, i.e., some ${\eta}^{\sigma}_{i}$ should be less than $\bar{\eta}^{\sigma}$ and others greater. This is motivated by the requirement to approximate the continuous distribution of the extra internal energy, which according to the equipartition theorem is given by $\dfrac{1}{2} m^{\sigma} {\bar{\eta}}^{\sigma2}=\dfrac{1}{2} I^{\sigma} {T^{\sigma}}$. A discrete set of ${\eta}^{\sigma}_{i}$ values spanning $\bar{\eta}^{\sigma}$ helps in achieving a more accurate representation of this energy.

\section{Verification and validation}\label{sec:Verification and validation}

For demonstrating the performance of the proposed DBM, five benchmark cases are considered. First, a multicomponent diffusion problem is simulated to demonstrate the DBM's capacity to handle interactions between distinct non-premixed species. Second, the Sod shock tube problem is adopted to verify the applicability of the DBM to compressible flows. Third, thermal Couette flow is carried out to validate the DBM’s adaptability to fluid systems where hydrodynamic behaviors and thermal TNE effects coexist. Fourth, the KH instability is calculated to illustrate the DBM's performance in depicting fluid flows with complicated interfacial evolutions. Finally, the RT instability further validates the capability of the DBM in simulating complex fluid behaviors under external force fields.

\subsection{Multicomponent diffusion}
In multicomponent systems, diffusion is a fundamental transport mechanism and the cornerstone of mass transfer, governing molecular mixing, interfacial dynamics, and the approach to thermodynamic equilibrium. Given its critical role in fluid dynamics, transport phenomena, and combustion, any reliable multicomponent DBM must first be validated against classical diffusion laws to ensure physical fidelity.

We simulate the multicomponent diffusion with the following initial conditions:
\[\begin{cases}\left(n^A,n^B,n^C\right)_L=\left(0.1,0.6,0.3\right) \text{,}\\\left(n^A,n^B,n^C\right)_R=\left(0.6,0.3,0.1\right) \text{,}&\end{cases}\]
where the symbol $L$ labels the left zone ($ -0.02\leq x\leq 0.00$), and $R$ indicates the right region ($ 0.00< x\leq 0.02$). Labels $A$, $B$, and $C$ are assigned to three kinds of independent chemical species, with the molar mass normalized to $m^{\sigma}=1$ for mathematical convenience. The velocity and temperature of this system are $\mathbf{u}=0$ and $T=1$. For the $x$-direction, second-order extrapolation is adopted to impose boundary constraints, whereas periodic boundary treatments are utilized along the $y$-direction. Based on Fick’s law of diffusion, the analytical formula for species concentrations under isothermal diffusion conditions is presented below~\cite{Bird2002AMR}:
\[
n^\sigma = \frac{n^{\sigma}_{L}+n^{\sigma}_{R}}{2}
- \frac{n^{\sigma}_{L}-n^{\sigma}_{R}}{2}
\;\mathrm{erf}\!\left(\frac{x}{\sqrt{4\Theta  t}} \right) \text{,}
\]
where $\mathrm{erf}$ represents the complementary error term, and $\Theta=0.001$ denotes the diffusivity.

As a preliminary step, a grid independence study is carried out to guarantee the credibility of numerical outcomes. To achieve this goal, four simulations are performed under different mesh configurations. Specifically, the horizontal grid resolution is set to
$N_x = 10$, $20$, $40$, and $80$, respectively, with the vertical grid consistently set to \(N_y = 1\). The time step $\Delta t = 1 \times 10^{-5}$, and the relaxation time $\tau^{A}=\tau^{B}= \tau^{C} =1.0\times 10^{-3}$.
\begin{figure}
	\begin{center}
		\includegraphics[width=0.6\textwidth]{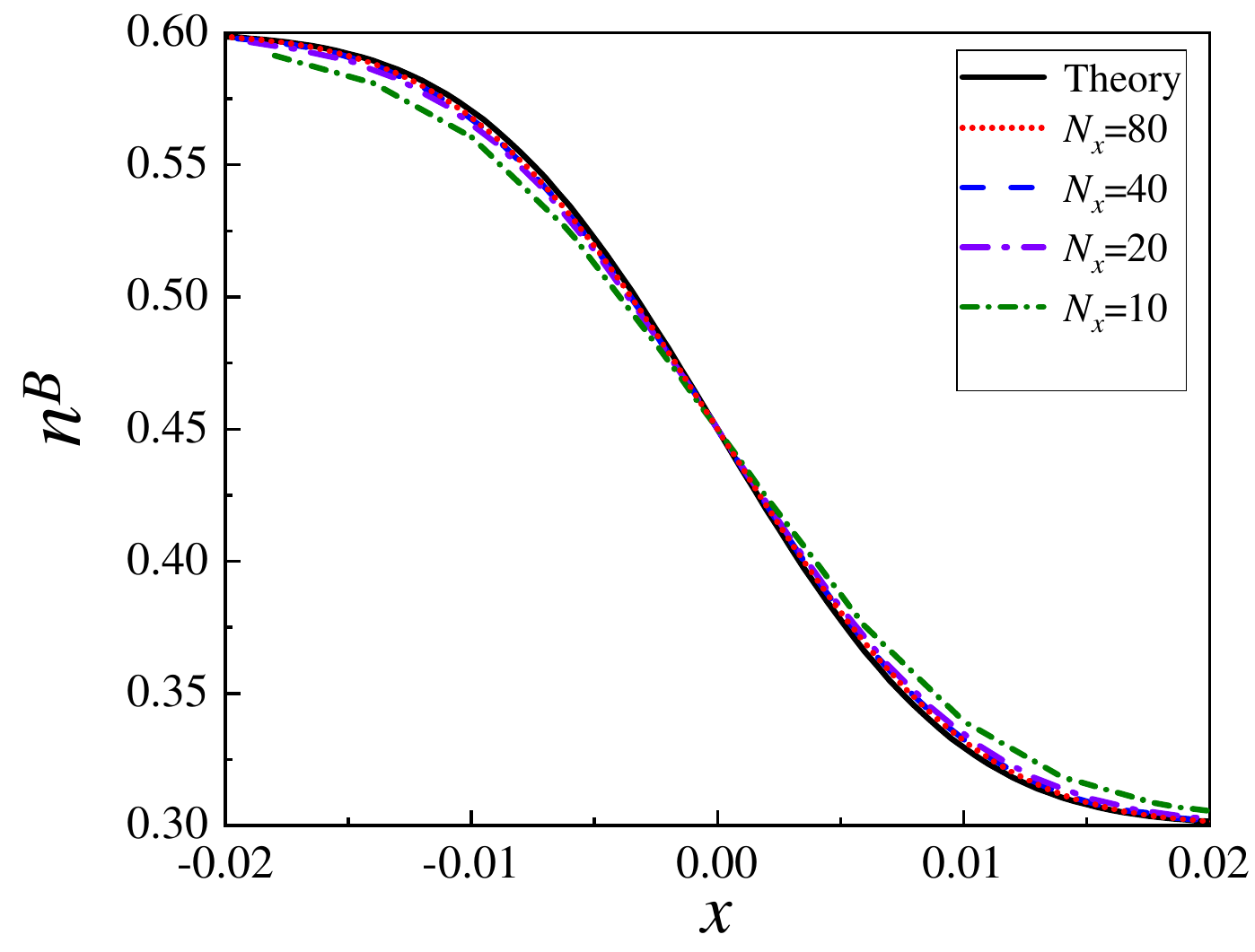}
	\end{center}
	\caption{Grid independence study: horizontal concentration profiles of species $B$ concentration at $t = 0.05$. The solid curves match the theoretical prediction, and dashed profiles illustrate numerical outputs obtained on various grid resolutions.}
	\label{Fig02}
\end{figure}

Figure \ref{Fig02} displays the concentration profile of species $B$. The green, purple, blue, and red dashed lines correspond to simulations with four different mesh resolutions, and the solid line stands for the theoretical solution. It is evident that the numerical solutions converge toward the theoretical solution as the grid density increases, with discrepancies between simulations of similar resolutions diminishing for finer meshes. Notably, the solutions for $N_x = 40$ and $N_x = 80$ are almost identical, and both are quite close to the theory prediction. To balance numerical accuracy and computational efficiency, a horizontal grid resolution of $N_x = 80$ is employed in the subsequent simulations.

\begin{figure}
	\begin{center}
		\includegraphics[width=0.6\textwidth]{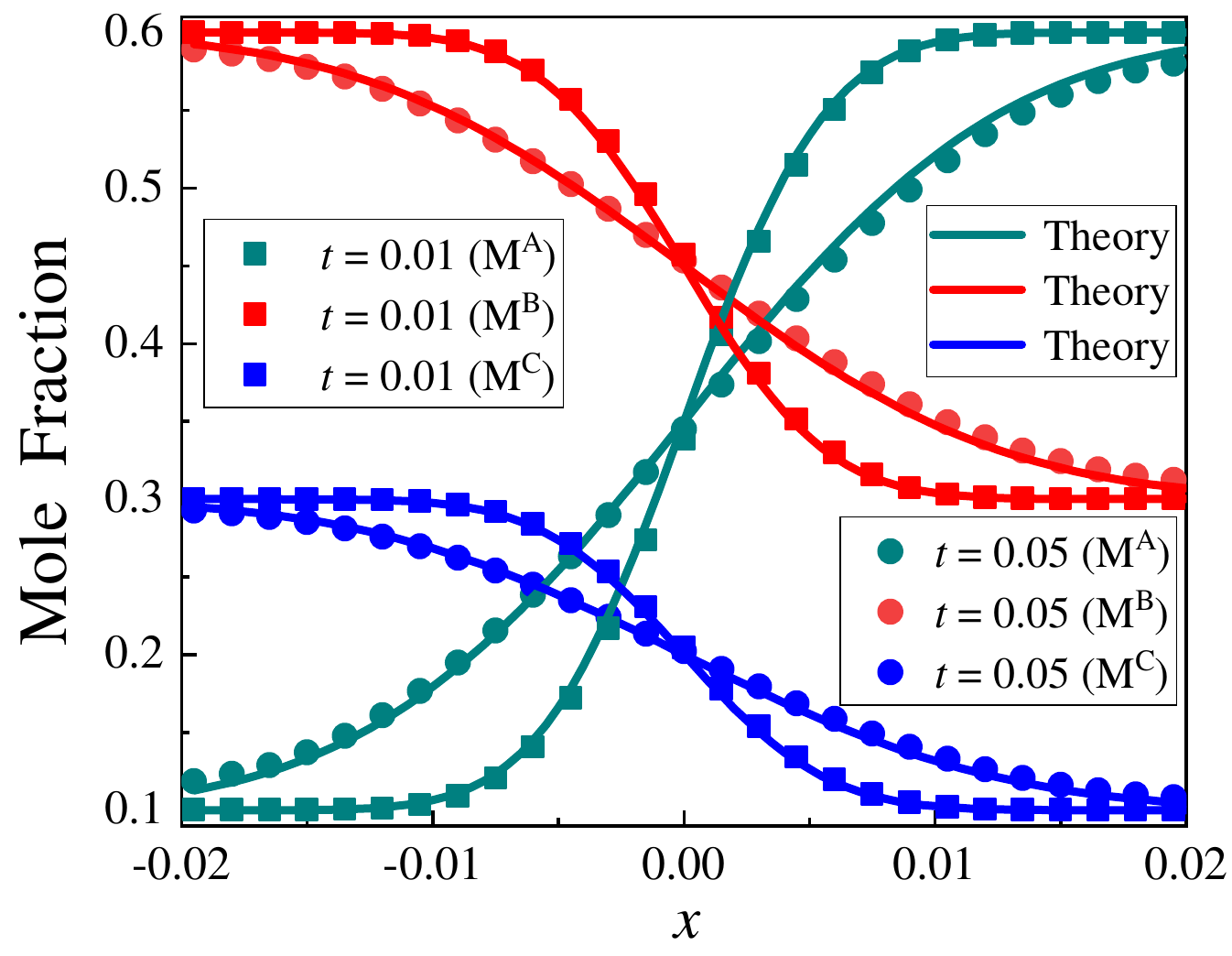}
	\end{center}
	\caption{Multicomponent diffusion process: concentration profiles of species $A$, $B$, and $C$ at $t = 0.01$ and $t = 0.05$. The symbols correspond to the numerical outcomes produced with the DBM simulations, and the solid lines denote the theoretical  solutions. }
	\label{Fig03}
\end{figure}

Figure \ref{Fig03} illustrates the concentration distributions of components $A$, $B$, and $C$ along the $x$-direction during the diffusion process.
Circles and squares correspond to the numerical results at $t = 0.01$ and $t = 0.05$, respectively, and the solid curves correspond to the theoretical solutions.
Green, red, and blue are used to distinguish components $A$, $B$, and $C$, respectively.
An obvious consistency can be observed between the DBM numerical outputs and theoretical solutions over the whole diffusion development process, confirming the capability of the DBM to faithfully capture multicomponent diffusion dynamics.

\subsection{Sod shock tube}
The Sod shock tube configuration serves as a standard test case for studying shock wave dynamics. The configuration consists of a sealed tube containing a compressible gas initially partitioned by a pressure discontinuity. Upon release, this discontinuity evolves into a flow featuring fundamental structures including a shock front, a rarefaction wave, and a contact discontinuity. The presence of these well-defined flow features makes the problem ideal for validating the accuracy and stability of numerical models. Unlike traditional single-component models, the present DBM is capable of simulating multicomponent systems with distinct molecular properties. To demonstrate its effectiveness for high-speed compressible flows, we compute the Sod shock tube with the following initial conditions:
\[\begin{cases}\left(n^A,n^B,n^C,p\right)_L=\left(1,0,0,1\right) \text{,}\\\left(n^A,n^B,n^C,p\right)_R=\left(0,0.05,0.075,0.1\right) \text{,}&\end{cases}\]
where the $L$ and $R$ correspond to the prescribed initial macroscopic quantities in the left zone spanning ($-1.0 \leq x \leq 0.0$) and the  right region covering  ($0.0 < x \leq 1.0$), separately. The two regions are initialized in a resting state, where the flow velocity satisfies $\mathbf{u}=0$. The molar masses of the three species are set as $m^A=m^B=m^C = 1$. Based on the relations $\rho^\sigma = m^\sigma n^\sigma$ and $T = p/\sum_\sigma n^\sigma$, the corresponding initial conditions yield $(\rho_L, \rho_R) = (1.0, 0.125)$ and $(T_L, T_R) = (1.0, 0.8)$. The grid is $N_x \times N_y = 2000 \times 1$, the spatial step $\Delta x = \Delta y = 1.0 \times 10^{-3}$, the time step $\Delta t = 1.0 \times 10^{-5}$, and the relaxation time  $\tau^{A}=\tau^{B}= \tau^{C} =1.0\times 10^{-5}$. Furthermore, the computational domain employs inflow and outflow boundaries in the $x$-direction, together with periodic boundaries in the $y$-direction.
\begin{figure}[!ht]
	\begin{center}
		\includegraphics[width=0.9\textwidth]{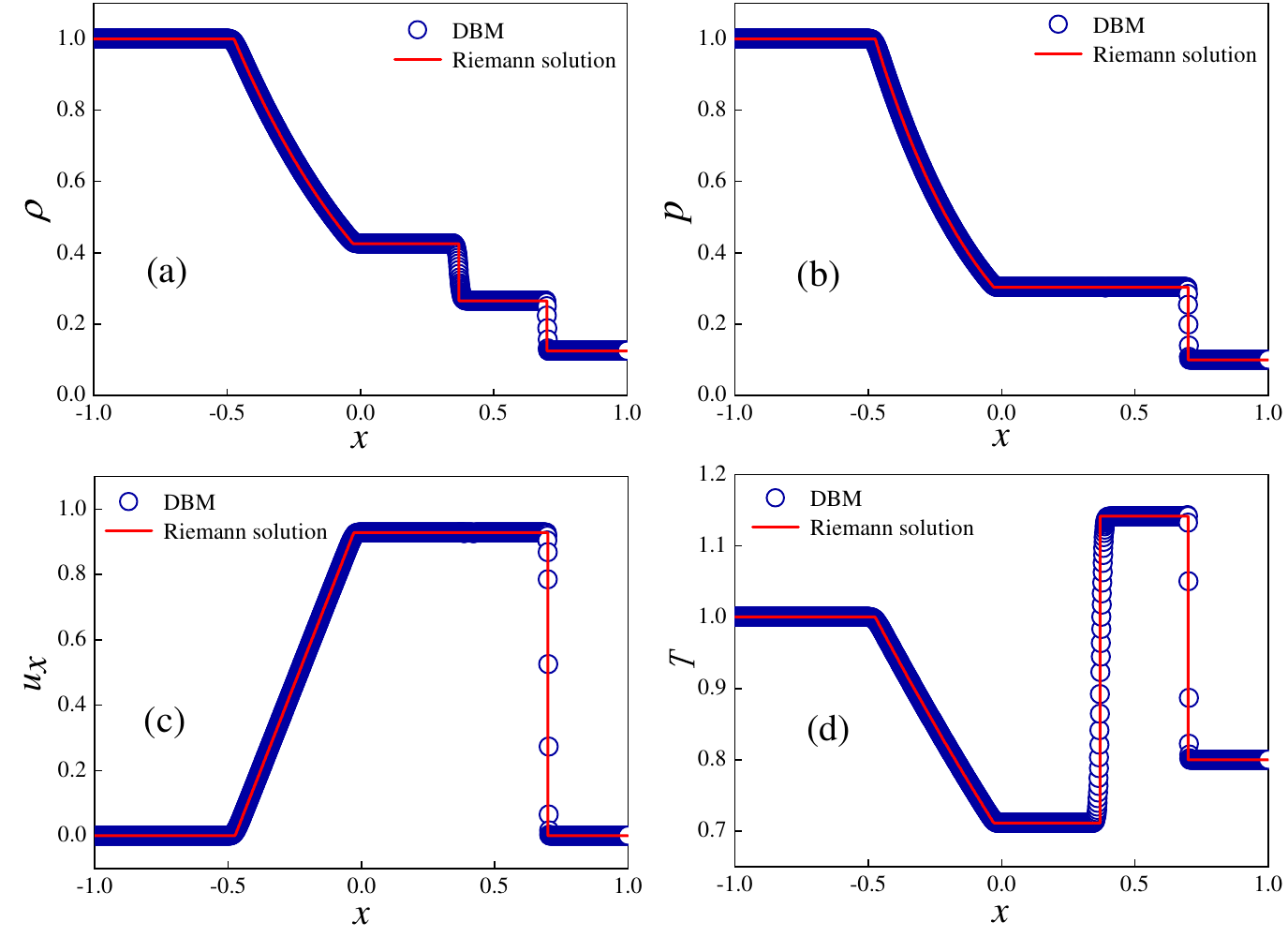}
	\end{center}
	\caption{Distributions of density (a), pressure (b), horizontal velocity (c), and temperature (d) for the Sod shock tube at $t=0.4$ }
	\label{Fi}
\end{figure}

Figure~\ref{Fi}(a) - (d) illustrate spatial distributions of density, pressure, x-direction velocity and temperature for the Sod shock tube when $t=0.4$. Discrete markers stand for numerical data from the DBM, while continuous solid curves correspond to Riemann solutions. During the evolution, three characteristic structures can be clearly identified: the rarefaction wave propagating leftward, the contact discontinuity associated with jumps in density and temperature, and the shock front on the right characterized by steep gradients. The close agreement between numerical and Riemann results demonstrates the capability of the DBM in accurately resolving discontinuities and capturing the essential wave patterns of the Sod tube problem.

\subsection{Thermal Couette flow}
Thermal Couette flow serves as a classic fluid dynamic case, featuring laminar flow trapped between two parallel plates that travel at separate velocities. In this flow, adjacent fluid layers slide relative to one another, resulting in a linear velocity profile across the channel. Due to these features, thermal Couette flow is widely employed as a benchmark for evaluating numerical methods applied to viscous and thermal flow problems. In this work, the simulation of thermal Couette flow is conducted with three objectives: to confirm the adaptability of the DBM across different values of the specific heat ratio, to illustrate its ability to capture nonequilibrium behaviors under hydrodynamic and thermodynamic constraints, and to validate the performance of the DBM when applied to premixed compressible fluid species.

The initial setup is as follows:
a premixed flow composed of species $A$, $B$, and $C$ is enclosed by two unbounded parallel plates that maintain a fixed gap height $H = 0.1$. The initial species concentrations are set as $(n^A, n^B, n^C) = (0.1, 0.3, 0.6)$, with uniform molar mass $m^{\sigma}=m_0 = 1$, temperature $T^{\sigma}=T_0 = 1$, and  initial velocity $\mathbf{u}=0$ across all species. The top plate slides tangentially with a fixed velocity $u_0=0.1$, while the bottom plate keeps static. Given the uniform flow distribution along the $x$-direction, this physical setup can be simplified into a one-dimensional model. The computational grid is configured with dimensions $ N_x \times N_y = 1 \times 200 $, the spatial step is set as $ \Delta x = \Delta y = 5.0 \times 10^{-4} $, and the relaxation time  $\tau^{A}=\tau^{B}= \tau^{C} =1.0\times 10^{-3}$. Additionally, a nonequilibrium extrapolation method is applied at the upper and lower walls, and periodic conditions are implemented along the lateral boundaries on two sides.

The theoretical expression for the horizontal velocity spatial variation takes the subsequent form~\cite{Batchelor2000Fluid,Watari2003PRE}.
\[u_x = \frac{y}{H} u_0 + \frac{2}{\pi} u_0 \sum_{n=1}^{\infty} \left[ \frac{(-1)^n}{n} \exp\left(-n^2 \frac{\pi^2 \mu t}{\rho H^2}\right) \sin\left( \frac{n \pi y}{H} \right) \right], \]
where $ \mu = {\tau}/{\rho T} $ indicates the dynamic viscosity.
Upon reaching steady-state conditions, the spatial temperature variation across the $y$-direction can be analytically expressed via the following expression~\cite{Batchelor2000Fluid,Watari2003PRE}:
$$
T = T_0 + \frac{u_0^2}{2c_p} \frac{y}{H} \left( 1 - \frac{y}{H} \right),
$$
where $c_p = {\gamma}/{(\gamma - 1)} $ denotes the isobaric specific heat and each species owns the same specific heat ratio $\gamma^{\sigma}= \gamma$.
\begin{figure}[!ht]
	\begin{center}
		\includegraphics[width=1\textwidth]{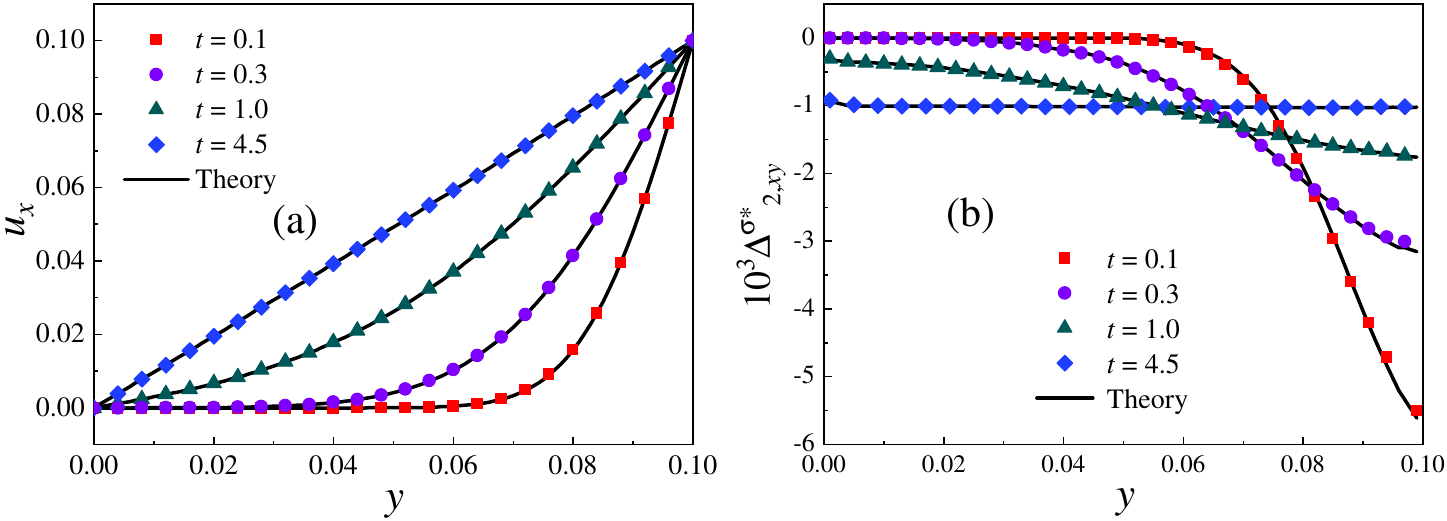}
	\end{center}
	\caption{Vertical distribution of the horizontal speed $u_x$ and $\Delta^{\sigma *}_{2,xy}$ under diverse time stages of Couette flow development. The symbols correspond to numerical outcomes of the DBM framework, and solid lines reflect the theoretical results.}
	\label{Fig05}
\end{figure}

Figure \ref{Fig05} displays the vertical distribution curves of the horizontal velocity $u_x$ (a) and the nonequilibrium quantities $\Delta^{\sigma *}_{2,xy}$ of species $\sigma=A$ (b) in the thermal Couette flow with a fixed specific heat ratio $\gamma=7/5$ at selected time instants $t=0.1$, $0.3$, $1.0$, and $4.5$, respectively. The discrete symbols correspond to results from the DBM, and the solid curves represent the corresponding theoretical solutions.
The theoretical solution of $\Delta^{\sigma *}_{2,xy}$ is obtainable through the Chapman--Enskog  analysis, and the specific expressions are given as follows:
\[
\begin{aligned}
	\Delta_{2,xy}^{\sigma*}
	=& -\frac{n^\sigma}{\tau^{\sigma}} \biggl[
	\bigl(u_x^\sigma - u_x\bigr)
	+ u_x \bigl(u_y^\sigma - u_y\bigr)
	+  \bigl(u_x^\sigma - u_x\bigr)\bigl(u_y^\sigma - u_y\bigr)
	\biggr] \\
	&-\frac{\rho^\sigma T^\sigma \tau^{\sigma}}{(m^\sigma)^2 }
	\Bigl(\partial_x u_y^\sigma + \partial_y u_x^\sigma\Bigr)
	+n^\sigma\bigl(u_x^\sigma - u_x\bigr)\bigl(u_y^\sigma - u_y\bigr).
\end{aligned}
\]
As depicted, the numerical results agree well with theoretical predictions across all measured time instants. These results verify that the DBM can faithfully capture fluid hydrodynamic evolutions and characterize relevant nonequilibrium behaviors.
\begin{figure}[!ht]
	\begin{center}
		\includegraphics[width=0.55\textwidth]{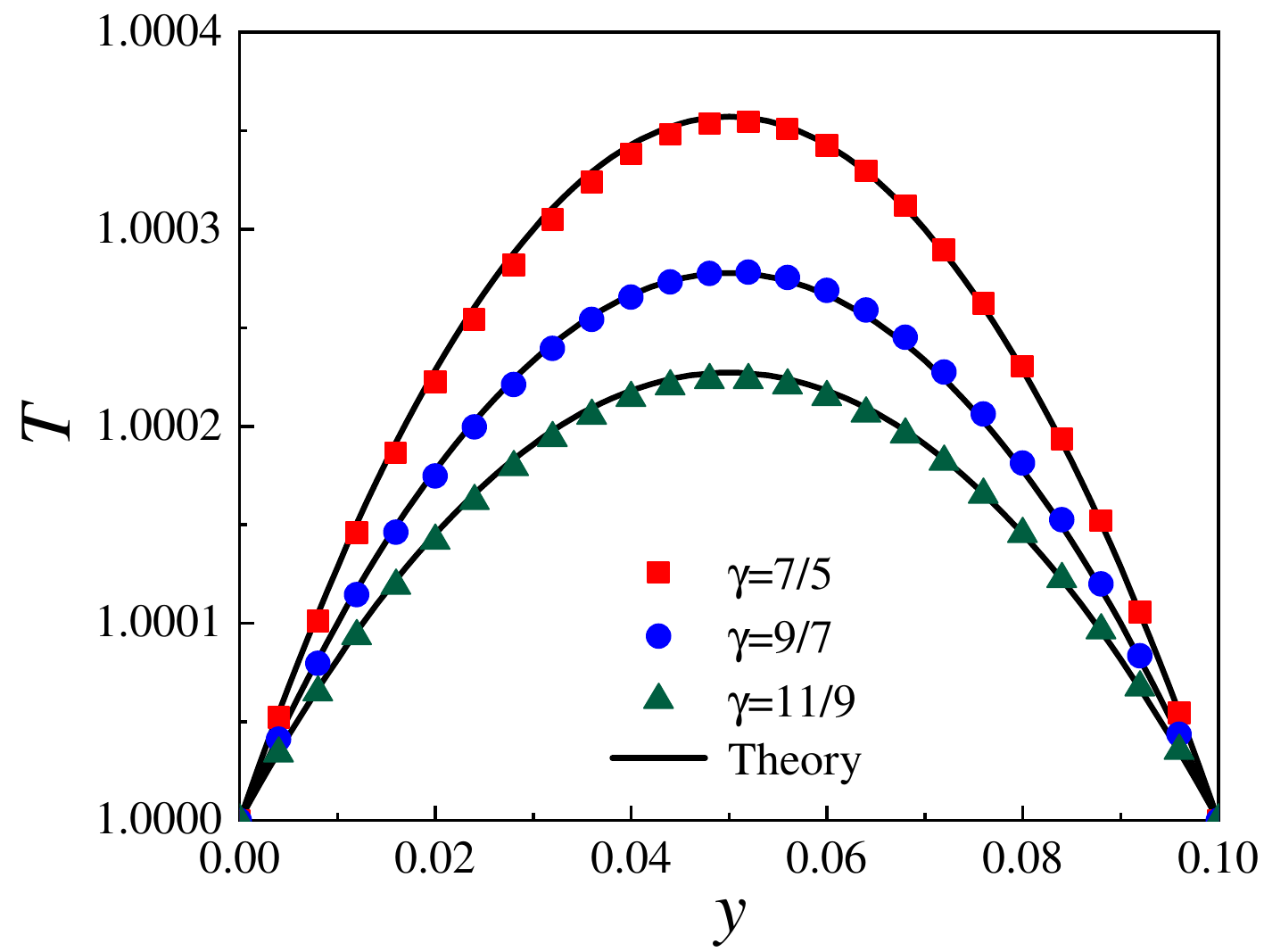}
	\end{center}
	\caption{Vertical spatial profiles of temperature within steady-state Couette flow under various specific heat ratios $\gamma=7/5$, $\gamma=9/7$, and $\gamma=11/9$, respectively.}
	\label{Fig06}
\end{figure}

Figure \ref{Fig06} displays the steady state temperature distributions in thermal Couette flow. Numerical results for various specific heat ratios of $\gamma = 7/5$, $9/7$, and $11/9$ are represented by triangles, squares, and circles, respectively, and the corresponding theoretical solutions are indicated by solid lines. Excellent agreement is observed between the simulated and theoretical profiles across all values of $\gamma$, confirming the capability of the DBM to accurately model flows with varying specific heat ratios.

\subsection{Kelvin--Helmholtz instability}
The KH instability, induced by velocity shear across fluid interfaces, is a typical interfacial instability commonly encountered in nature phenomena and engineering applications. To validate the capability of the present DBM in capturing interfacial dynamics and complex flow evolution, the numerical simulation of the KH instability is performed.
\begin{figure}[!ht]
	\begin{center}
		\includegraphics[width=0.35\textwidth]{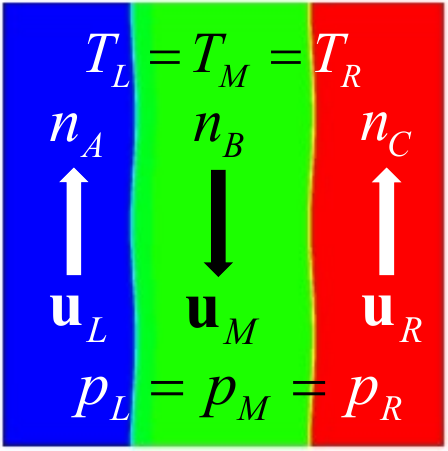}
	\end{center}
	\caption{Original computational layout of the KH instability employed in the present simulations.}
	\label{F8}
\end{figure}

Figure \ref{F8} illustrates the original computational layout of the three-component fluid system. A square calculation domain with identical horizontal and vertical dimensions $(L_x = L_{y} = 1)$ is established, which falls into three segmented regions along the $x$-direction: $0 < x \leq x_1$, $x_1 < x \leq x_2$, and $x_2 < x \leq L_x$.
The interface separating the left and central subdomains is placed at $x_1 = 0.3L_x$, while the boundary between the central and right segments is set at $x_2 = 0.7L_x$. The initial configuration consists of species $A$, $B$, and $C$ occupying the left, central, and right subdomains in sequence. The flow velocity is set as $\mathbf{u} = u_x {\mathbf{e}_x}+u_y {\mathbf{e}_y}$, with $u_x=0$. Densities and temperatures are uniform ($n_L = n_M = n_R$, $T_L = T_M = T_R$), thereby	ensuring uniform pressure ($p_L = p_M = p_R$) across the interfaces according to $p^\sigma = n^\sigma T^\sigma$.
To initiate the roll-up dynamics of the KH instability, a cosine-shaped perturbation is imposed on the two interfaces:
\[ w=w_{0}\cos\left( \frac{4\pi y}{L_y} \right),  \]
where $w_0=L_x/200$ is the amplitude. To maintain smoothness at the interfaces, the concentrations and velocities are initially set with continuous profiles:
\[
n^{A} = \frac{n_{L}}{2} - \frac{n_{L}}{2} \tanh\left(\frac{x-x_{1}+w}{W_n} \right) \text{,}
\]
\[
n^{C} = \frac{n_{R}}{2} - \frac{n_{R}}{2} \tanh\left(\frac{x-x_{2}+w}{W_n} \right) \text{,}
\]
\[
n^{B} = n_{M}-n^{A}-n^{C} \text{,}
\]
and
\[{u_y}=\left\{\begin{aligned}
	& \frac{u_{L}+u_{M}}{2} - \frac{u_{L}-u_{M}}{2}  \tanh\left(\frac{x-x_{1}+w}{W_u} \right), \quad 0 <x \leq \frac{L_x}{2}, \\
	& \frac{u_{M}+u_{R}}{2} - \frac{u_{M}-u_{R}}{2}  \tanh\left(\frac{x-x_{2}+w}{W_u} \right), \quad \frac{L_x}{2} <x \leq L_{x},
\end{aligned}\right.\]
where $W_{n}$ and $W_{u}$ correspond to the widths of the concentration and velocity transition layers in sequence, and $(u_{L},u_{M},u_{R})=(0.5,-0.5,0.5)$.

Furthermore, the mesh grid is $N_x \times N_y = 500 \times 500 $, the spatial step $ \Delta x = \Delta y = 1.0\times 10^{-3} $, the time step $ \Delta t = 2.0 \times 10^{-5} $, the relaxation time  $\tau^{A}=\tau^{B}=\tau^{C} =2.0\times 10^{-4}$, and the molar masses are $m^A=1$, $m^B=1.5$, and $m^C=2$ in sequence. Additionally, two types of boundary constraints are implemented as follows: the specular reflection scheme for the $x$-direction, and periodic conditions across the $y$-direction.
\begin{figure}[!ht]
	\begin{center}
		\includegraphics[width=1.0\textwidth]{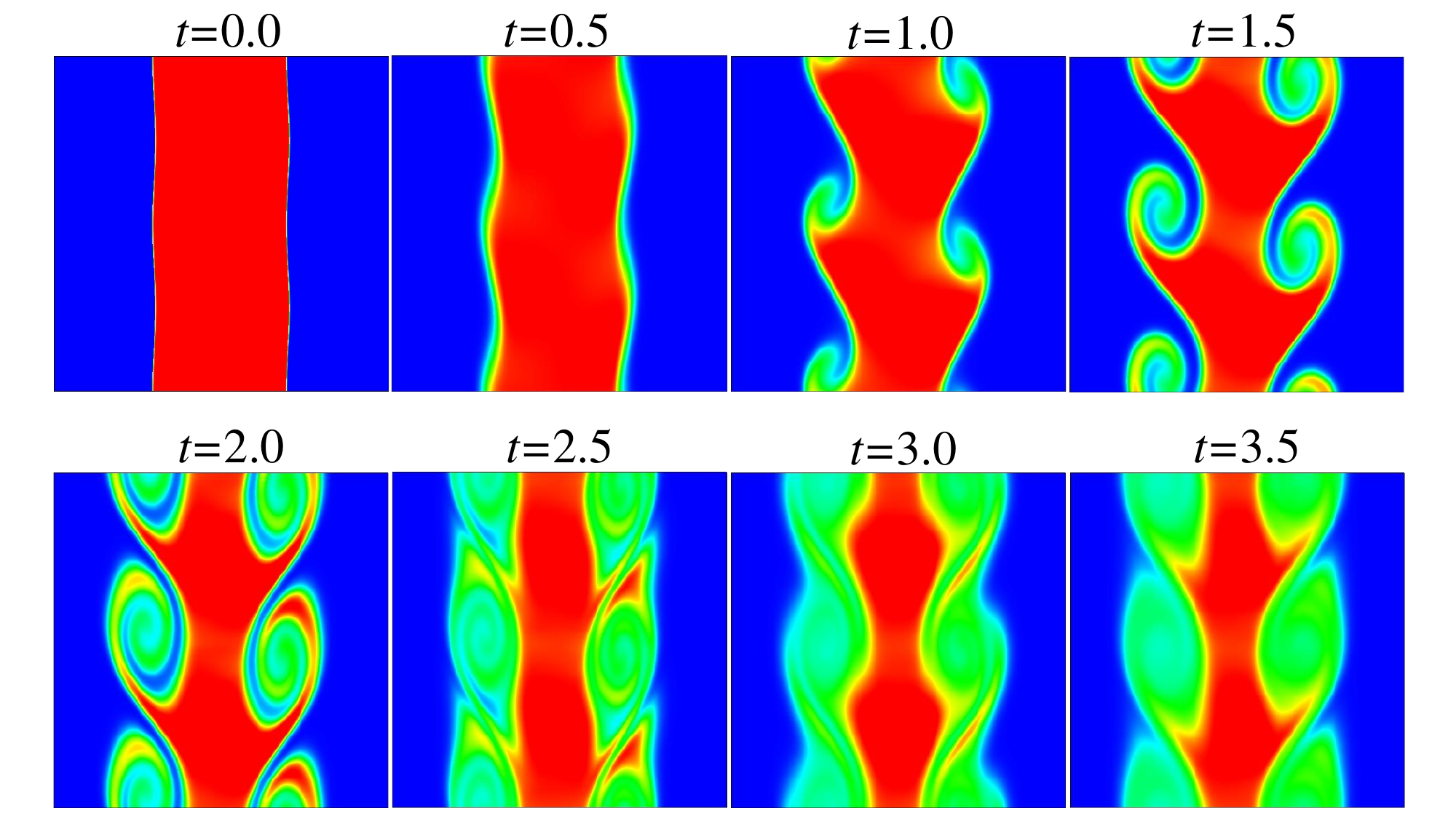}
	\end{center}
	\caption{Evolution of the molar fraction of species $B$ during the KH instability at various time instants. The contour scale ranges from blue ($0$) to red ($1$). }
	\label{F99}
\end{figure}

Figure \ref{F99} depicts the spatial contours of species $B$ throughout the development of miscible KH instability at several representative time instants. Initially, the interface exhibits small-amplitude perturbations; as the system evolves, the shear flow amplifies these disturbances, and the concentration transition layer broadens due to molecular diffusion. The interface progressively deforms, transitioning from a smooth, regular shape to a more irregular and complex configuration. Midway through the evolution, pairs of vortices form, with the central fluid penetrating into neighboring regions. At later times, the vortices grow continuously, leading to the emergence of billow-like structures and increasingly convoluted interfaces. Eventually, the fluid motion becomes chaotic, and the KH instability drives substantial mixing between the diverse fluid species. Overall, the observed dynamics are in good agreement with previously reported KH instability scenarios~\cite{Lin2024CTP}, demonstrating that the proposed model well characterizes the nonlinear interfacial evolution processes.

\subsection{Rayleigh--Taylor instability}
Rayleigh--Taylor (RT) instability is a typical multi-material interfacial instability. When the density gradient opposes the effective pressure gradient, small perturbations at the fluid interface amplify continuously and evolve over time. Compared with single-component flows, RT instability in multicomponent media exhibits far richer and more complex dynamics. It is governed not only by external body forces such as gravity, but also by critical physical effects: interspecies mass diffusion, intercomponent viscosity contrasts, and cross-interface mass transport. Under a constant external force field, mechanical equilibrium is destabilized when a denser fluid overlays a lighter one. Interfacial perturbations grow progressively, eventually forming nonlinear, interpenetrating ``bubble-spike'' structures, alongside interpenetration and dynamic mixing of distinct components. To validate the ability of the multicomponent DBM to simulate multi-material interfacial instabilities, species mixing, and complex flow dynamics under external forcing, we perform numerical calculations of compressible RT instability in this part.

The calculation region for the two-component RT simulation is a rectangle with length $L_{x}=0.05$ and height $L_{y}=0.2$. The computational domain consists of two individual regions separated by a central perturbed interface. Component $A$ occupies the upper region with  a particle mass $m^A=2.0$, while component $B$ is distributed in the lower region with $m^B=1.0$. A gravitational field \(\boldsymbol{a}=(0,a_y)\) is imposed on the entire system, with $a_{y}=-2.0$. In the static equilibrium state with $\nabla p=\rho \boldsymbol{a}$, the initial concentrations are configured as:
\[\left\{\begin{aligned}
	&n^A = \frac{p_m}{T_u} \exp \left[ \frac{-a_{y}m^A}{T_u} \left( y_m - y \right) \right], \quad n^B=0,\quad y > y_m, \\
	&n^B = \frac{p_m}{T_d} \exp \left[ \frac{-a_{y}m^B}{T_u} \left( y_m - y \right) \right], \quad n^A=0, \quad y < y_m,
\end{aligned}\right.\]
where $p_m=4.0$ is defined as the pressure located on the material interface, and the position of the perturbed interface is expressed as $y_m = L_y/2 + L_y/50\cos(\pi x/L_x)$. For thermal initial conditions, the entire computational domain is initialized with a uniform temperature field: $T=1.0$.
Furthermore, the mesh grid is $N_x \times N_y = 256 \times 1024 $,  the time step $ \Delta t =  2.5\times10^{-6} $, and the relaxation time  $\tau^{\sigma}=3.5\times 10^{-5}$. Furthermore, the specular reflection boundary condition is imposed along the $x$- and $y$-direction boundaries.
\begin{figure}[!ht]
	\begin{center}
		\includegraphics[width=0.6\textwidth]{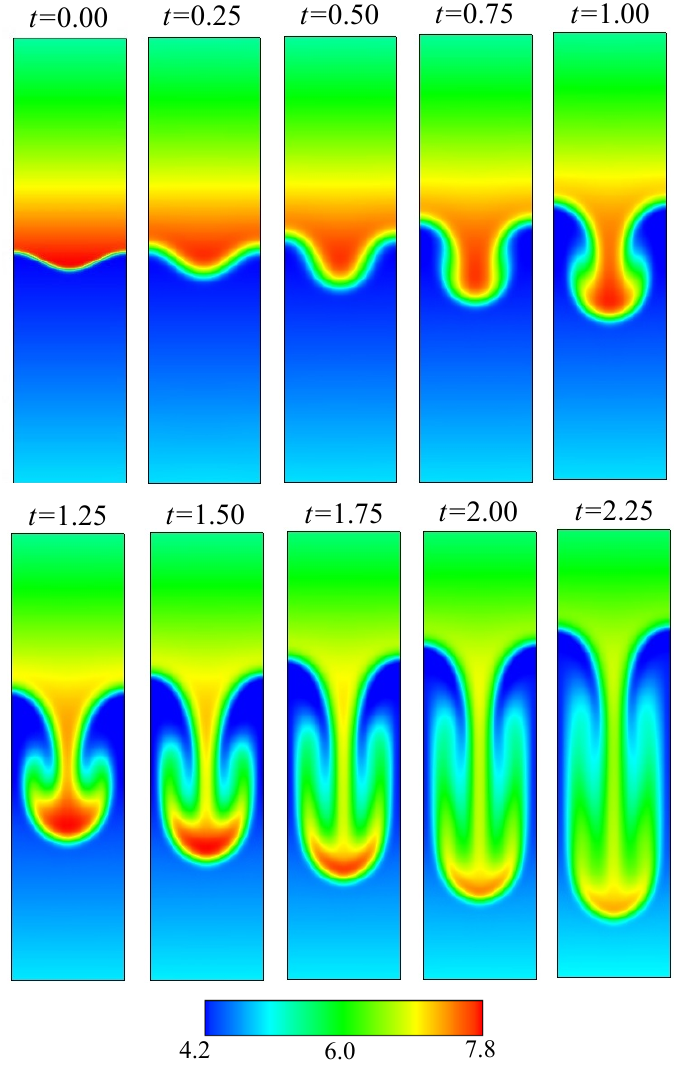}
	\end{center}
	\caption{Density contours during the RT instability development for component $B$. }
	\label{F09}
\end{figure}

Figure \ref{F09} illustrates the evolution of density contours for the RT instability of component $B$ at successive time instants: $t=0.00$, $0.25$, $0.50$, $0.75$, $1.00$, $1.25$, $1.50$, $1.75$, $2.00$, and $2.25$, respectively. At the start of the simulation ($t=0.00$), the flow field is in hydrostatic equilibrium. The interface separating the light and dense fluids bears a predefined small-amplitude cosine perturbation, accompanied by a smooth density distribution and distinct stratification. For $0.25 \le t \le 0.50$, the interfacial perturbation begins to evolve slowly. Diffusive effects slightly broaden the density transition layer; the perturbation amplitude increases exponentially, while the interface retains a regular sinusoidal profile. After $t=0.75$, the instability enters a rapid growth phase. Dense fluid descends to form spike structures, whereas the light fluid rises to develop bubble-like configurations, inducing severe asymmetric deformation of the fluid interface. 
Within $1.00 \le t \le 1.50$, intensified interfacial shear excites the KH instability, which gradually rolls up the spike tails and yields characteristic mushroom-shaped structures. At late times ($1.75 \le t \le 2.25$), fluid mixing is substantially enhanced. The mushroom structures undergo further elongation, and the density field exhibits multi-scale turbulent mixing features. The entire evolutionary process agrees well with the classical growth regime of RT instability. The above numerical results fully verify the reliability of the DBM for simulating fluid systems featuring complex structures under external body forces.

\section{Conclusions}\label{sec:conclusion}
In this work, a high-order Burnett-level multicomponent discrete Boltzmann model (DBM) is developed for compressible multicomponent flows under external body forces, which constitutes a substantial extension of our previously proposed single-component forced DBM framework. The model adopts a well-designed D2V25 discrete velocity set with excellent spatial isotropy and numerical stability. Both the discrete equilibrium distribution function and external force term are rigorously constructed via a precise moment-matching strategy, which satisfies 25 independent kinetic moment relations and ensures rigorous consistency between mesoscopic kinetic descriptions and macroscopic hydrodynamic governing laws. Theoretical analysis demonstrates that the proposed model can exactly recover the Burnett-level governing equations of forced compressible multicomponent flows in the continuum limit. Beyond macroscopic hydrodynamic behaviors, the present DBM inherently captures interspecies mass diffusion and coupled multicomponent thermodynamic nonequilibrium (TNE) effects, which are inaccessible to conventional Navier--Stokes-level models and single-component DBMs.

Five canonical benchmark cases are systematically performed to fully validate the physical reliability, numerical accuracy, and robustness of the developed model. The multicomponent diffusion test verifies the model’s capability of resolving mutual mass transport and species mixing behaviors between different fluid components. The Sod shock tube case confirms the model’s competence in capturing typical discontinuities and high-gradient flow structures in compressible flows. The thermal Couette flow simulation validates the model’s ability to simultaneously reproduce hydrodynamic evolution and TNE responses under temperature-gradient conditions. The Kelvin--Helmholtz instability benchmark demonstrates the model’s robustness in resolving intricate interfacial deformation, mixing, and instability development in multicomponent fluid mixtures. The Rayleigh--Taylor instability test further verifies the model’s effectiveness in simulating external-force-driven multicomponent interfacial flows and complex flow field evolution. Collectively, all numerical validations illustrate that the proposed Burnett-level multicomponent DBM provides a reliable and versatile numerical tool for modeling forced compressible multicomponent nonequilibrium flows.

This study fills the research gap in high-order kinetic modeling for compressible multicomponent systems with external forces, and effectively extends the application scope of high-precision Burnett-level DBMs from single-component to practical multi-fluid flow scenarios. In future work, the model can be further extended to reactive multicomponent flows, turbulent mixing problems, and multiscale nonequilibrium flow systems to broaden its applicability in complex engineering and physical flow problems.

\begin{acknowledgments}
This work is supported by National Natural Science Foundation of China (under Grant Nos. U2242214, 12572341), Natural Science Foundation of Fujian Province (Grant No. 2026J001415), Guangdong Basic and Applied Basic Research Foundation (Grant No. 2024A1515010927), Humanities and Social Science Foundation of the Ministry of Education in China (Grant No. 24YJCZH163). This work is supported by the opening project of State Key Laboratory of Explosion Science and Safety Protection (Beijing Institute of Technology). The opening project number is KFJJ26-17M.
\end{acknowledgments}

\section*{Data Availability}
The data that support the findings of this study are available from the corresponding author upon reasonable request.

\appendix

\section{}\label{A}
For the square matrix $\mathbf{C}$ shown in Eq.~\ref{Matrix_form_feq}, its specific expression is given below.
\begin{equation}
	\mathbf{C}
	=
	\left(
	\begin{array}{cccc}
		{C_{11}} & C_{12} & \cdots  & C_{1N}  \\
		{C_{21}} & C_{22} & \cdots  & C_{2N}  \\
		\vdots  & \vdots  & \ddots  & \vdots   \\
		{C_{N1}} & C_{N2} & \cdots  & C_{NN}
	\end{array}
	\right)
	\text{,}
\end{equation}
with elements $C_{1i}=1$, $C_{2i}=v_{ix}^{\sigma}$, $C_{3i}=v_{iy}^{\sigma}$, $C_{4i}={v_{i}^{\sigma 2}}+{\eta_{i}^{\sigma 2}}$, $C_{5i}={v_{ix}^{\sigma 2}}$,
$C_{6i}=v_{ix}^{\sigma}v_{iy}^{\sigma}$, ${{C}_{7i}}={v_{iy}^{\sigma 2}}$, $C_{8i}=\left( {v_{i}^{\sigma 2}}+{\eta _{i}^{\sigma 2}} \right)v_{ix}^{\sigma}$, $C_{9i}=\left( {v_{i}^{\sigma 2}}+{\eta _{i}^{\sigma 2}} \right)v_{iy}^{\sigma}$,
$C_{10i}={v_{ix}^{\sigma 3}}$, $C_{11i}={v_{ix}^{\sigma 2}}v_{iy}^{\sigma}$, $C_{12i}=v_{ix}^{\sigma}{v_{iy}^{\sigma 2}}$, $C_{13i}=v_{iy}^{\sigma 3}$,
$C_{14i}=\left( {v_{i}^{\sigma 2}}+{\eta _{i}^{\sigma 2}} \right){v_{ix}^{\sigma 2}}$, $C_{15i}=\left( {v_{i}^{\sigma 2}}+{\eta _{i}^{\sigma 2}} \right)v_{ix}^{\sigma}v_{iy}^{\sigma}$, $C_{16i}=\left({v_{i}^{\sigma 2}} +{\eta _{i}^{\sigma 2}} \right) {v_{iy}^{\sigma 2}}$, $C_{17i}={v_{ix}^{\sigma4}}$, $C_{18i}={v_{ix}^{\sigma 3}v_{iy}^{\sigma}}$, $C_{19i}={v_{ix}^{\sigma 2}v_{iy}^{\sigma 2}}$, $C_{20i}={v_{ix}^{\sigma}v_{iy}^{\sigma 3}}$, $C_{21i}={v_{iy}^{\sigma4}}$, $C_{22i}=\left({v_{i}^{\sigma 2}} +{\eta _{i}^{\sigma 2}} \right) {v_{ix}^{\sigma 3}}$, $C_{23i}=\left({v_{i}^{\sigma 2}} +{\eta _{i}^{\sigma 2}} \right) {v_{ix}^{\sigma 2}v_{iy}^{\sigma}}$, $C_{24i}=\left({v_{i}^{\sigma 2}} +{\eta _{i}^{\sigma 2}} \right) {v_{ix}^{\sigma}v_{iy}^{\sigma 2}}$, $C_{25i}=\left({v_{i}^{\sigma 2}} +{\eta _{i}^{\sigma 2}} \right) {v_{iy}^{\sigma 3}}$. The inverse matrix $\mathbf{C}^{-1}$ can be computed using MATLAB.

The column matrix of equilibrium kinetic moments ${\mathbf{{\hat f}^{\sigma\rm{eq}}}}$ in Eq. (\ref{Matrix_form_feq}) takes the form,
\begin{equation}
	{\mathbf{{\hat f}^{\sigma\rm{eq}}}}=[\hat f_{1}^{\sigma\rm{eq}}, \hat f_{2}^{\sigma\rm{eq}}, \dots, \hat f_{N}^{\sigma\rm{eq}}]^{\rm{T}}
	\label{Mo}
	\text{,}
\end{equation}
with elements $\hat f_{1}^{\sigma \rm{eq}}=n^{\sigma}$, $\hat f_{2}^{\sigma \rm{eq}}=n^{\sigma}u_{x}$, $\hat f_{3}^{\sigma \rm{eq}}=n^{\sigma}u_{y}$, $\hat f_{4}^{\sigma \rm{eq}}=n^{\sigma}\biggl[\bigl(D+I^{\sigma}\bigr)T/m^{\sigma}+u^{2}\biggr]$, $\hat f_{5}^{\sigma \rm{eq}}=n^{\sigma}\Big(T/m^{\sigma}+u_{x}^{2}\Big)$, $\hat f_{6}^{\sigma \rm{eq}}=n^{\sigma}u_{x}u_{y}$, $\hat f_{7}^{\sigma \rm{eq}}=n^{\sigma}\Big(T/m^{\sigma}+u_{y}^{2}\Big)$, $\hat f_{8}^{\sigma \rm{eq}}=n^{\sigma}u_{x}\biggl[\bigl(D+I^{\sigma}+2\bigr)T/m^{\sigma}+u^{2}\biggr]$, $\hat f_{9}^{\sigma \rm{eq}}=n^{\sigma}u_{y}\biggl[\bigl(D+I^{\sigma}+2\bigr)T/m^{\sigma}+u^{2}\biggr]$, $\hat f_{10}^{\sigma \rm{eq}}=3n^{\sigma}u_{x}T/m^{\sigma}+n^{\sigma}u_{x}^{3}$, $\hat f_{11}^{\sigma \rm{eq}}=n^{\sigma}u_{y}T/m^{\sigma}+n^{\sigma}u_{x}^{2}u_{y}$, $\hat f_{12}^{\sigma \rm{eq}}=n^{\sigma}u_{x}T/m^{\sigma}+n^{\sigma}u_{y}^{2}u_{x}$, $\hat f_{13}^{\sigma \rm{eq}}=3n^{\sigma}u_{y}T/m^{\sigma}+n^{\sigma}u_{y}^{3}$, $\hat f_{14}^{\sigma \rm{eq}}=n^{\sigma}\biggl[\bigl(D+I^{\sigma}+2\bigr)T/m^{\sigma}+u^{2}\biggr]+n^{\sigma}u_{x}^{2}\biggl[\bigl(D+I^{\sigma}+4\bigr)T/m^{\sigma}+u^{2}\biggr]$, $\hat f_{15}^{\sigma \rm{eq}}=n^{\sigma}u_{x}u_{y}\biggl[\bigl(D+I^{\sigma}+4\bigr)T/m^{\sigma}+u^{2}\biggr]$, $\hat f_{16}^{\sigma \rm{eq}}=n^{\sigma}\biggl[\bigl(D+I^{\sigma}+2\bigr)T/m^{\sigma}+u^{2}\biggr]+n^{\sigma}u_{y}^{2}\biggl[\bigl(D+I^{\sigma}+4\bigr)T/m^{\sigma}+u^{2}\biggr]$, $\hat f_{17}^{\sigma \rm{eq}}=6n^{\sigma}u_{x}^{2}T/m^{\sigma}+n^{\sigma}u_{x}^{4}+3n^{\sigma}T^{2}/m^{2 \sigma}$, $\hat f_{18}^{\sigma \rm{eq}}=3n^{\sigma}u_{x}u_{y}T/m^{\sigma}+n^{\sigma}u_{x}^{3}u_{y}$, $\hat f_{19}^{\sigma \rm{eq}}=n^{\sigma}\bigl(u_{x}^{2}+u_{y}^{2}\bigr)T/m^{\sigma}+n^{\sigma}u_{y}^{2}u_{x}^{2}+n^{\sigma}T^{2}/m^{2 \sigma}$, $\hat f_{20}^{\sigma \rm{eq}}=3n^{\sigma}u_{x}u_{y}T/m^{\sigma}+n^{\sigma}u_{y}^{3}u_{x}$, $\hat f_{21}^{\sigma \rm{eq}}=6n^{\sigma}u_{y}^{2}T/m^{\sigma}+n^{\sigma}u_{y}^{4}+3n^{\sigma}T^{2}/m^{2 \sigma}$, $\hat f_{22}^{\sigma \rm{eq}}=3n^{\sigma}u_{x}T/m^{\sigma}\biggl[\bigl(D+I^{\sigma}+4\bigr)T/m^{\sigma}+u^{2}\biggr]+n^{\sigma}u_{x}^{3}\biggl[\bigl(D+I^{\sigma}+6\bigr)T/m^{\sigma}+u^{2}\biggr]$, $\hat f_{23}^{\sigma \rm{eq}}=n^{\sigma}u_{y}T/m^{\sigma}\biggl[\bigl(D+I^{\sigma}+4\bigr)T/m^{\sigma}+u^{2}\biggr]+n^{\sigma}u_{y}u_{x}^{2}\biggl[\bigl(D+I^{\sigma}+6\bigr)T/m^{\sigma}+u^{2}\biggr]$, $\hat f_{24}^{\sigma \rm{eq}}=n^{\sigma}u_{x}T/m^{\sigma}\biggl[\bigl(D+I^{\sigma}+4\bigr)T/m^{\sigma}+u^{2}\biggr]+n^{\sigma}u_{x}u_{y}^{2}\biggl[\bigl(D+I^{\sigma}+6\bigr)T/m^{\sigma}+u^{2}\biggr]$, $\hat f_{25}^{\sigma \rm{eq}}=3n^{\sigma}u_{y}T/m^{\sigma}\biggl[\bigl(D+I^{\sigma}+4\bigr)T/m^{\sigma}+u^{2}\biggr]+n^{\sigma}u_{y}^{3}\biggl[\bigl(D+I^{\sigma}+6\bigr)T/m^{\sigma}+u^{2}\biggr]$.

The column matrix of the force term ${\mathbf{{\hat F^{\sigma}}}}$ in Eq. (\ref{Matrix_form_force}) takes the form,
\begin{equation}
	{\mathbf{{\hat F^{\sigma}}}}=[\hat F_{1}^{\sigma},\hat F_{2}^{\sigma},\dots,\hat F_{N}^{\sigma}]^{\rm{T}}
	\text{,}
\end{equation}
with elements $\hat F_{1}=0$, $\hat F_{2}=n^{\sigma}a_{x}$, $\hat F_{3}=n^{\sigma}a_{y}$, $\hat F_{4}=2n^{\sigma}u_{x}^{\sigma}a_{x}+2n^{\sigma}u_{y}^{\sigma}a_{y}$, $\hat F_{5}=2n^{\sigma}u_{x}^{\sigma}a_{x}$, $\hat F_{6}=2n^{\sigma}u_{x}^{\sigma}a_{y}+2n^{\sigma}u_{y}^{\sigma}a_{x}$, $\hat F_{7}=2n^{\sigma}u_{y}^{\sigma}a_{y}$, $\hat F_{8}=3n^{\sigma}a_{x}u_{x}^{\sigma 2}+n^{\sigma}a_{x}u_{y}^{\sigma 2}+n^{\sigma}a_{x}\bigl(D+I^{\sigma }+2\bigr)T/m^{\sigma }+2n^{\sigma}a_{y}u_{x}^{\sigma}u_{y}^{\sigma}$, $\hat F_{9}=2n^{\sigma}a_{x}u_{x}^{\sigma}u_{y}^{\sigma}+3n^{\sigma}a_{y}u_{y}^{\sigma 2}+n^{\sigma}a_{y}u_{x}^{\sigma 2}+n^{\sigma}a_{y}\bigl(D+I^{\sigma }+2\bigr)T/m^{\sigma }$, $\hat F_{10}=3n^{\sigma}a_{x}\bigl(u_{x}^{\sigma 2}+T/m^{\sigma }\bigr)$, $\hat F_{11}=2n^{\sigma}a_{x}u_{x}^{\sigma}u_{y}^{\sigma}+n^{\sigma}a_{y}\bigl(u_{x}^{\sigma 2}+T/m^{\sigma }\bigr)$, $\hat F_{12}=n^{\sigma}a_{x}\bigl(u_{y}^{\sigma 2}+T/m^{\sigma }\bigr)+2n^{\sigma}a_{y}u_{x}^{\sigma}u_{y}^{\sigma}$, $\hat F_{13}= 3n^{\sigma}a_{y}\bigl(u_{y}^{\sigma 2}+T/m^{\sigma }\bigr)$, $\hat F_{14}=2n^{\sigma}u_{x}^{\sigma}a_{x}\biggl[2u_{x}^{\sigma 2}+u_{y}^{\sigma 2}+\bigl(D+I^{\sigma }+5\bigr)T/m^{\sigma }\biggr]+2n^{\sigma}a_{y}u_{y}^{\sigma}\bigl(u_{x}^{\sigma 2}+T/m^{\sigma }\bigr)$, $\hat F_{15}=n^{\sigma}u_{y}^{\sigma}a_{x}\biggl[3u_{x}^{\sigma 2}+u_{y}^{\sigma 2}+\bigl(D+I^{\sigma }+4\bigr)T/m^{\sigma }\biggr]+n^{\sigma}a_{y}u_{x}^{\sigma}\biggl[u_{x}^{\sigma 2}+3u_{y}^{\sigma 2}+\bigl(D+I^{\sigma }+4\bigr)T/m^{\sigma }\biggr]$, $\hat F_{16}=2n^{\sigma}u_{y}^{\sigma}a_{y}\biggl[2u_{y}^{\sigma 2}+u_{x}^{\sigma 2}+\bigl(D+I^{\sigma }+5\bigr)T/m^{\sigma }\biggr]+2n^{\sigma}a_{x}u_{x}^{\sigma }\bigl(u_{y}^{\sigma 2}+T/m^{\sigma }\bigr)$, $\hat F_{17}=4n^{\sigma}a_{x}u_{x}^{\sigma }\bigl(u_{x}^{\sigma 2}+T/m^{\sigma }\bigr)$, $\hat F_{18}=3n^{\sigma}a_{x}u_{x}^{\sigma 2}u_{y}+n^{\sigma}a_{y}u_{x}^{\sigma 3}+3n^{\sigma}u_{y}^{\sigma}a_{x}T/m^{\sigma }+3n^{\sigma}u_{x}^{\sigma}a_{y}T/m^{\sigma }$, $\hat F_{19}=2n^{\sigma}a_{y}u_{y}^{\sigma}\bigl(u_{x}^{\sigma 2}+T/m^{\sigma }\bigr)+2n^{\sigma}a_{x}u_{x}^{\sigma}\bigl(u_{y}^{\sigma 2}+T/m^{\sigma }\bigr)$, $\hat F_{20}=3n^{\sigma}a_{y}u_{y}^{\sigma 2}u_{x}^{\sigma}+n^{\sigma}a_{x}u_{y}^{\sigma 3}+3n^{\sigma}u_{x}^{\sigma}a_{y}T/m^{\sigma }+3n^{\sigma}u_{y}^{\sigma}a_{x}T/m^{\sigma }$, $\hat F_{21}=4n^{\sigma}a_{y}u_{y}^{\sigma}\bigl(u_{y}^{\sigma 2}+T/m^{\sigma }\bigr)$, $\hat F_{22}=5n^{\sigma}a_{x}u_{x}^{\sigma 4}+3n^{\sigma}a_{x}u_{x}^{\sigma 2}\bigl(D+I^{\sigma }+9\bigr)T/m^{\sigma }+3n^{\sigma}a_{x}\biggl[u_{y}^{\sigma 2}+\bigl(D+I^{\sigma }+4\bigr)T/m^{\sigma }\biggr]T/m^{\sigma }+3n^{\sigma}a_{x}u_{x}^{\sigma 2}u_{y}^{\sigma 2}+2n^{\sigma}a_{y}u_{x}u_{y}\bigl(u_{x}^{\sigma 2}+3T/m^{\sigma }\bigr)$, $\hat F_{23}=n^{\sigma}a_{y}u_{x}^{\sigma 4}+4n^{\sigma}a_{x}u_{x}^{\sigma 3}u_{y}^{\sigma}+n^{\sigma}a_{y}u_{x}^{\sigma 2}\biggl[3u_{y}^{\sigma 2}+\bigl(D+I^{\sigma }+7\bigr)T/m^{\sigma }\biggr]+2n^{\sigma}a_{x}u_{x}u_{y}^{\sigma}\biggl[u_{y}^{\sigma 2}+\bigl(D+I^{\sigma }+7\bigr)T/m^{\sigma }\biggr]+n^{\sigma}a_{y}\biggl[3u_{y}^{\sigma 2}+\bigl(D+I^{\sigma }+4\bigr)T^{\sigma 2}/m^{\sigma 2}\biggr]$, $\hat F_{24}=n^{\sigma}a_{x}u_{y}^{\sigma 4}+4n^{\sigma}a_{y}u_{y}^{\sigma 3}u_{x}^{\sigma}+n^{\sigma}a_{x}u_{y}^{\sigma 2}\biggl[3u_{x}^{\sigma 2}+\bigl(D+I^{\sigma }+7\bigr)T/m^{\sigma }\biggr]+2n^{\sigma}a_{y}u_{x}^{\sigma}u_{y}^{\sigma}\biggl[u_{x}^{\sigma 2}+\bigl(D+I^{\sigma }+7\bigr)T/m^{\sigma }\biggr]+n^{\sigma}a_{x}\biggl[3u_{x}^{\sigma 2}+\bigl(D+I^{\sigma }+4\bigr)T^{\sigma 2}/m^{\sigma 2}\biggr]$, $\hat F_{25}=5n^{\sigma}a_{y}u_{y}^{\sigma 4}+3n^{\sigma}a_{y}u_{y}^{\sigma 2}\bigl(D+I^{\sigma }+9\bigr)T/m^{\sigma }+3n^{\sigma}a_{y}\biggl[u_{x}^{\sigma 2}+\bigl(D+I^{\sigma }+4\bigr)T/m^{\sigma }\biggr]T/m^{\sigma }+3n^{\sigma}a_{y}u_{y}^{\sigma 2}u_{x}^{\sigma 2}+2n^{\sigma}a_{x}u_{y}^{\sigma}u_{x}^{\sigma}\bigl(u_{y}^{\sigma 2}+3T/m^{\sigma }\bigr)$.

\nocite{*}
\bibliography{References}

\end{document}